\documentclass[journal=acsnano,manuscript=article]{achemso}

\usepackage{graphicx}% Include figure files
\usepackage{multirow}
\usepackage{threeparttable}
\usepackage{dcolumn}% Align table columns on decimal point
\usepackage{bm}% bold math
\usepackage{mathtools,amssymb}
\usepackage{upgreek}
\usepackage{color}
\usepackage{float}
\usepackage[utf8]{inputenc}
\usepackage{mathptmx}
\usepackage{etoolbox}
\usepackage[symbol]{footmisc}
\usepackage{chemformula} % Formula subscripts using \ch{}
\usepackage[T1]{fontenc} % Use modern font encodings
\usepackage{amsmath}
\newcommand{\angstrom}{\text{\normalfont\AA}}
\makeatletter

\author{Vineet Pandey}
\affiliation[Indian Institute of Technology, Kharagpur]
{Materials Science Centre, Indian Institute of Technology, Kharagpur, West Bengal - 721302,India }
\author{Subhendu Mishra}
\affiliation[Indian Institute of Science]
{Materials Research Centre, Indian Institute of Science, Bengaluru - 560012, India}
\author{Nikhilesh Maity}
\affiliation[Indian Institute of Science]
{Materials Research Centre, Indian Institute of Science, Bengaluru - 560012, India}

\author{Sourav Paul}
\affiliation[Indian Institute of Technology, Kharagpur]
{Materials Science Centre, Indian Institute of Technology, Kharagpur, West Bengal - 721302,India }

\author{Abhijith M B}
\affiliation[Indian Institute of Technology, Kharagpur]
{Materials Science Centre, Indian Institute of Technology, Kharagpur, West Bengal - 721302,India }

\author{Ajit Roy}
\affiliation[Air Force Research Laboratory]
{Air Force Research Laboratory, Wright-Patterson Air Force Base, OH - 45433, United States}

\author{Nicholas R Glavin}
\affiliation[Air Force Research Laboratory]
{Air Force Research Laboratory, Wright-Patterson Air Force Base, OH - 45433, United States}

\author{Kenji Watanabe}
\affiliation[NIMS]
{Research Center for Electronic and Optical Materials, National Institute for Materials Science, 1-1 Namiki, Tsukuba 305-0044, Japan}

\author{Takashi Taniguchi}
\affiliation[NIMS]
{Research Center for Materials Nanoarchitectonics, National Institute for Materials Science, 1-1 Namiki, Tsukuba 305-0044, Japan}

\author{Abhishek Kumar Singh}
\email{abhishek@iisc.ac.in}
\affiliation[Indian Institute of Science]
{Materials Research Centre, Indian Institute of Science, Bengaluru - 560012, India}

\author{Vidya Kochat}
\email{vidya@matsc.iitkgp.ac.in}
\affiliation[Indian Institute of Technology, Kharagpur]
{Materials Science Centre, Indian Institute of Technology, Kharagpur, West Bengal - 721302,India }

\title[Probing interlayer interactions and commensurate-incommensurate transition in twisted bilayer graphene through Raman spectroscopy ]
  {Probing interlayer interactions and commensurate-incommensurate transition in twisted bilayer graphene through Raman spectroscopy}

\begin{document}

%\begin{tocentry}
%
%Some journals require a graphical entry for the Table of Contents.
%This should be laid out ``print ready'' so that the sizing of the
%Inside the \texttt{tocentry} environment, the font used is Helvetica
%8\,pt, as required by \emph{Journal of the American Chemical
%The surrounding frame is 9\,cm by 3.5\,cm, which is the maximum
%permitted for  \emph{Journal of the American Chemical Society}
%graphical table of content entries. The box will not resize if the
%content is too big: instead it will overflow the edge of the box.
%
%This box and the associated title will always be printed on a
%separate page at the end of the document.
%
%\end{tocentry}

\begin{abstract}
Twisted 2D layered materials have garnered a lot of attention recently as a class of 2D materials whose interlayer interactions and electronic properties are dictated by the relative rotation / twist angle between the adjacent layers. In this work, we explore a prototype of such a twisted 2D system, artificially stacked twisted bilayer graphene (TBLG), where we probe the changes in the interlayer interactions and electron-phonon scattering pathways as the twist angle is varied from 0 to 30$^{\circ}$, using Raman spectroscopy. The long range Moir\'e potential of the superlattice gives rise to additional intravalley and intervalley scattering of the electrons in TBLG which have been investigated through their Raman signatures. The density functional theory (DFT) calculations of the electronic band structure of the TBLG superlattices was found to be in agreement with the resonant Raman excitations across the van Hove singularities in the valence and conduction bands predicted for TBLG due to hybridization of bands from the two layers. We also observe that the relative rotation between the graphene layers has a marked influence on the second order overtone and combination Raman modes signalling a commensurate-incommensurate transition in TBLG as the twist angle increases. This serves as a convenient and rapid characterization tool to determine the degree of commensurability in TBLG systems.

\end{abstract}

%\keywords{\textbf{Keywords} twisted bilayer graphene, Raman spectroscopy, combination modes, density functional theory, commensurate-incommensurate transition}


2D layered materials with twist dependent stacking have been a hot topic of research recently, as the electronic properties of such systems critically depend on the superlattice formation due to relative rotation between the layers. A prototype system, twisted bilayer graphene (TBLG) has been a typical example of the superlattice-induced modified bandstructure with several new phases such as correlated insulator, superconductivity and ferromagnetism emerging at critical twist angles called the magic angles.
\cite{MATBLG_superconductivity,MATBLG_correlated_insulator,TBLG_ferromagnetism,TBLG_AHE} In a TBLG Moir\'e superlattice,  alternating AA and AB stacked regions emerge, leading to a hexagonal superlattice potential  which folds the bands into a mini Brillouin zone. The adjacent Dirac cones in this mini Brillouin zone hybridize leading to bands with reduced Fermi velocity near charge neutrality point, furthermore developing into flat bands for the critical angles close to 1.1$^{\circ}$ or the magic angles.\cite{Moirebands_PNAS} For small values of twist angles, the atoms adjust with the superlattice landscape forming an energetically favourable commensurate system, by gaining van der Waals energy, but at the expense of elastic energy. As the Moir\'e period becomes smaller, the van der Waals energy gain over the elastic energy loss is no longer compensated, leading to an incommensurate system of bilayers, where each layer is independent of the other.\cite{vdW_elastic_energy_TBLG} This shows that the mechanical and electronic degrees of freedom are closely related in TBLG systems, wherein the twist angle determines the modifications to the bandstructure and also the interlayer coupling. 

The 2D quantum materials-based technology projections are focussed on identifying artificially stacked 2D heterostructures with combined and novel functionalities of various 2D materials. In this context, it becomes very important to understand how this unique degree of freedom of twisting layers determines the fundamental properties  such as mechanical coupling, elasticity, electronic band structure and electron-phonon coupling in such 2D heterostructure stacks. In this work, we have investigated these properties in twisted graphene bilayers, which is a very promising system of homobilayers exhibiting a rich variation of the electronic properties with changing relative rotation between the layers. While at low angles, emergent phases attributed to flat bands have been observed, at high angles close to 30$^{\circ}$, TBLG forms an incommensurate superlattice with quasicrystalline order and mirrored Dirac cones originating from strong interlayer interaction.\cite{QCgraphene_Science,QCgraphene_PNAS,QCgraphene_Nanoletters} A significant experimental effort has gone into 
studies of the Moir\'e superlattices in TBLG concentrating on STM and ARPES. \cite{STM_1,STM_2,STM_3,STM_4,STM_5,STM_6,STM_7,STM_8,STM_9,ARPES_1,ARPES_2} We have studied the effect of the Moir\'e potential on the interlayer coupling and bandstructure by using Raman spectroscopy, which is a well-established powerful characterization technique to investigate the layer numbers, disorder and stacking orientations in bi and tri-layer graphene systems.\cite{Ferrari_Raman,Raman_disorder_1,Raman_disorder_2,Raman_stacking,Raman_misorientedBLG,Raman_TLG} Raman studies on folded graphene and chemical vapor deposition (CVD) grown rotationally stacked bilayer graphene showed emergence of new modes due to the intravalley and intervalley scattering of electrons by Moir\'e potential.\cite{Folded_Gr_1,Folded_Gr_3,Folded_Gr_4,CVD_TBLG_carbon,CVD_TBLG_ACSNano,CVD_TBLG_nanophotonics,CVD_TBLG_lowfreqmodes,CVD_TBLG_SSC,CVD_TBLG_JPCC,CVD_TBLG_JRamanSpec} Also it was found that Raman studies as a function of laser energy and twist angle can trace the positions of the van Hove singularities (vHs) in the conduction and valence bands of TBLG originating from the hybridization of the Dirac cones of the twisted graphene layers.\cite{Folded_Gr_2,CVD_TBLG_angleresolved_Raman,AS_TBLG_PRL} Modifications of the electronic band structure and the phonon dispersion in TBLG was suggested to give rise to frequency shifts of the 2D peak and well as the lineshape and width.\cite{Folded_Gr_red_vF,CVD_TBLG_APL,CVD_TBLG_SciRep,CVD_TBLG_Natcomm,CVD_TBLG_Nanoresearch,CVD_TBLG_SciRep_doping} On the other hand, the Raman studies in artificially stacked bilayer graphene systems has been very limited. Similar observations were noted in twisted CVD bilayers fabricated by PMMA-assisted tansfer, while in artificially stacked exfoliated bilayer graphene, studies reported the variations in in-plane anisotropy and superlattice-induced transverse acoustic Raman modes with the mis-orientation angle.\cite{AS_TBLG_2Dmater,AS_TBLG_inplane_anisotropy} For TBLG close to magic angles, the G and 2D bands were significantly influenced by the electron-phonon interactions and the areal coverage of AB and BA stacked regions separated by strain soliton regions.\cite{AS_TBLG_nearMA}

In this work, we have studied the electron-phonon scattering mechanisms in artificially stacked TBLG in detail and substantiated our results with that of folded and CVD grown graphene Raman signatures. The various intravalley and intervalley scattering mechanisms induced by Moir\'e potential have been investigated and the results have been explained by variations in the electronic bandstructure and density of states (DOS) obtained from density functional theory (DFT) calculations. In addition we observe that the resonance arising from the transition between the vHs in the valence and conduction band also presents interesting features in G and M bands. Our work shows that the tunability of the vHs in rotationally misoriented bilayer graphene can give rise to enhanced absorption leading to wavelength selective photodetectors in future. Finally we also explore the overtone and combination modes in TBLG which serves as a definitive signature of commensurate to incommensurate transition at higher twist angles and mechanical decoupling of the layers. 

\section{Results and discussion}

The TBLG samples were fabricated using the tear and stack method using a 2D transfer system (from CryoNano Labs) for various twist angles as shown in the optical micrographs in top panels of Fig.~1a-d.\cite{tear_and_stack}  The middle panel shows the superlattice evolution for the corresponding angles from which it is evident that at low angles ($<12.5^{\circ}$) a long-range Moir\'e pattern evolves and the TBLG can be approximately treated as a system with Moir\'e period governed translational symmetry. At larger twist angles ($>20^{\circ}$), the Moir\'e period is in the order of atomic length scale evolving to a quasi-crystalline (QC) state at twist angles of $30^{\circ}$. The FFT of these TBLG systems gives a direct estimation of the Moir\'e wavevector,  $\textbf{q}$, whose magnitude is plotted in Fig.~1e. The QC TBLG lattice which lacks translational symmetry shows a 12-fold rotational order in the FFT analysis. A geometrical analysis in the reciprocal space reveals that the additional periodicity due to the Moir\'e potential results in a new mini-Brillouin zone with a twist angle ($\theta$) dependent lattice vector which is the resultant of the wavevectors $\textbf{q}_1$ and $\textbf{q}_2$ corresponding to the top and bottom layers and is given by 
\begin{equation}
  q(\theta)=\frac{8\pi}{\sqrt{3}{a}}\sin\left(\frac{\theta}{2}\right)
\end{equation}
where $a=$~0.246~nm is the lattice parameter of graphene.\cite{Folded_Gr_4} The values obtained for $q(\theta)$ from this calculation perfectly matches the values obtained directly from the FFT analysis as can be seen from Fig.~1e.

The TBLG samples were studied through Raman spectroscopy using 532~nm excitation and the spectra corresponding to the twist angles $8^{\circ}$, $10^{\circ}$, $12.5^{\circ}$, $20^{\circ}$ and $30^{\circ}$ are shown in Fig.~2a,b along with a single layer graphene (SLG) and Bernal-stacked bilayer graphene (BLG). The G-peak arising from the doubly degenerate zone center $E_{2g}$ mode occurs at 1582~cm$^{-1}$ and shows negligible frequency shift with the twist angle. Interestingly at twist angle of $12.5^{\circ}$, we observe a huge enhancement of the G-peak intensity. Such an enhancement has been reported earlier in CVD grown bilayer graphene samples at misorientation angles of 	10$^{\circ}$ to 12$^{\circ}$ under 633 ~nm and 532~nm excitation.\cite{CVD_TBLG_carbon,CVD_TBLG_APL,AS_TBLG_PRL} This has been attributed to the resonant condition when the laser excitation energy matches the parallel band singularity in TBLG. Here there are a large number of states with the same energy difference, giving rise to Raman scattering paths with identical phases that interfere constructively and result in considerable enhancement of G-peak.\cite{CVD_TBLG_SciRep_doping,CVD_TBLG_angleresolved_Raman} When the Dirac cones of the top and bottom layers overlap, the DOS is modified due to the interactions leading to hybridization gap and vHs emerge in the conduction and valence bands of TBLG. When the excitation laser energy $E_{L}$ matches this energy separation between the conduction and valence band vHs, which in turn is dependent on the twist angle $\theta$, resonance occurs as per the condition
\begin{equation}
  \theta_C=\frac{3aE_{L}}{{\hbar}v_f4\pi}
\end{equation}
where $v_f$ is the Fermi velocity in SLG ($10^6$~m/s) and $\theta_C$ is the critical twist angle. For excitation wavelength of 532~nm (2.33~eV), $\theta_C\sim12.3^{\circ}$ which matches well with the twist angle where the resonance is observed experimentally. This is also the first observation of G-peak resonance in TBLG fabricated by tear and stack method. In the Raman map in Fig.~2c (given in red, bottom panel), the G-peak resonance is clearly observable in the TBLG region, whereas the region of SLG has much lower intensity. We show later that this picture of parallel bands is in accordance with the band structure obtained from DFT simulations as well.

As opposed to the G-peak, the 2D peak shows significant variations in frequency and intensity as a function of twist angle. At twist angles larger than $10^{\circ}$, we observe a notable blue shift in our TBLG samples. Previous studies have attributed this blue shift to reduction in Fermi velocity in TBLG.~\cite{Folded_Gr_red_vF,AS_TBLG_PRL} The maximum blue shift occurs at $\theta_C$ and decreases at lower and higher twist angles as seen in the plot in Fig.~2e. But theoretical calculations show that significant reduction in Fermi velocity happens for twist angles $<5^{\circ}$, contrary to our observations.\cite{DFT9_symmetry_breaking,DFT10_atomic_corrugation,breakdown_TBLG_DFT11} The possible origin of the 2D peak shift will be discussed later. Another interesting aspect of the 2D peak is the single Lorentzian behaviour at twist angles larger than $8^{\circ}$. This is in contrast with the convoluted peak picture for the Bernal-stacked BLG and TBLG with misorientation angles less than $5^{\circ}$ reported earlier.\cite{Ferrari_Raman,AS_TBLG_nearMA} It proves that the Dirac nature of the bands from individual layers are retained at higher twist angles. With increase in twist angle, the intensity ratio of the 2D peak to the G peak, $I_{2D}/I_G$ increases to about 4 times that of SLG, while the FWHM of the single Lorenztian reduces and becomes similar to that of SLG as shown in Fig.~2d. This can be attributed to the differences in the scattering mechanisms as a function of twist angle described in the schematics in Fig.~2b. For $\theta>\theta_C$, the separation between the vHs, $\Delta{E_{vHs}}>E_L$ and hence the energy dispersion is linear with negligible interaction between the layers. The iTO phonons contribute to the double resonance process of intralayer intervalley scattering between the K and K$^{\prime}$ valleys of the two layers. Since the frequencies of the scattering phonons, energy dispersion and electron-phonon coupling are all identical for the two layers, the resulting 2D peak should have a higher intensity than SLG and also similar FWHM. At $\theta=\theta_C$, $\Delta{E_{vHs}}\approx E_L$, and a large number of optical transitions occur due to the vHs which increases the joint DOS. There is also a transition from linear dispersion to a flat band dispersion, due to which the phonons giving rise to intervalley scattering have larger wavevectors which also explains the huge blue shift of the 2D peak at $\theta_C$. The flat band and the joint DOS also gives rise to slight variations in the wavevectors of the scattering phonons leading to an increased FWHM. Finally, at $\theta<\theta_C$, $\Delta{E_{vHs}}< E_L$, due to the increased interlayer interactions, along with intralayer scattering, interlayer intervalley scattering is also possible. This gives rise to a complex interference between the scattered phonons reducing the intensity and increasing the FWHM at low twist angles as observed in Fig.~2d. The 2D Raman maps for the various twist angles are shown in Fig.~2c, which clearly indicate that the intensity is uniform throughout the entire TBLG region and also is a good experimental proof for a spatially homogeneous superlattice structure.  

At low angles of $8^{\circ}$, we observe an additional peak at 1625~cm$^{-1}$, which is absent for higher twist angles as shown in Fig.~2f. This peak is attributed to the R$^{\prime}$-peak which arises due to the intra-valley double resonance process, where the excited electron is elastically scattered by its interaction with the Moir\'e potential through the rotational wavevector, $q(\theta)$ to the point in the same valley with equal and opposite momentum. A phonon with a wavevector same as $q(\theta)$ given by $Q_{intra}$ is created and scatters the electron back inelastically, which finally combines with a hole to give the Raman shift observed. From the phonon dispersion curve of graphene, we observe that $Q_{intra}$ corresponds to the LO phonon frequency of 1625~cm$^{-1}$ which lies close to the $\Gamma$ point in the first Brillouin zone along the $\Gamma$K direction and has strong electron-phonon coupling.~\cite{Folded_Gr_4,Dresselhaus_phonondispersion} For this type of intra-valley scattering for a twist angle of $\theta$ the incident photon energy is given by 
\begin{equation}
  E_L^{intra}(\theta)=\hbar{v_F}q(\theta)
\end{equation}
For $ {E_L}^{intra}=2.33$~eV, $q(\theta)$ turns out to be 0.365~${\angstrom}^{-1}$ corresponding to a twist angle of $7.1^{\circ}$ from Fig.~1e, which is very close to the experimental twist angle of $8^{\circ}$.

%DFT Explanaton starts here

The experimental Raman signatures have also been corroborated by first-principles DFT calculations performed using Vienna ab initio simulation package (VASP).\cite{DFT7_ab_initio,DFT8_ab_initio} The electronic band structure of SLG has a linear band dispersion in the vicinity of high-symmetry point K and results in four-fold degenerate states from the $p_{z}$ orbital of the carbon atom. On the other hand, in TBLG, at energies far from the Fermi level, the bands are modified and spilt, resulting in the appearance of extrema for some of the sub-bands close to the M-point in the Brillouin zone as indicated by 1, 2, 3 and 4 in Fig. 3b.\cite{DFT16_AA_AB_bandstructure} These saddle points near M point in the band structure of TBLG results in twist-induced spikes (singularity 2 and 3) and twist induce kinks (singularity 1 and 4) in DOS known as vHs shown in Fig. 3e. These saddle points/vHs play an important role in determining the optical transitions when light is incident on these systems. These vHs differences ($\Delta E_{vHs}$) give the information of strength and the energy of the transitions. In TBLG, the allowed transitions are from 1-3 and 2-4, while 1-4 and 2-3 transitions are forbidden.\cite{DFT17_ramanbands_kinks_spikes} The $\Delta E_{vHs}$ variation with twist angle of TBLG for allowed and forbidden transitions are given in Fig. 3f. From the Fig. 3f, it is observed that the vHs splitting increases with increasing the twist angle of TBLG, which is consistent with the previous literature.\cite{DFT14_unraveling_vHs_TBLG} When the excitation laser energy $E_{L}$ matches the energy separation between the conduction and valence band vHs of allowed transition (transition 1-3 or 2-4), a large number of excitations occur. The calculated allowed transition energies of TBLG for twist angles $9.43^{\circ}$, $13.2^{\circ}$, $21.8^{\circ}$, and $27.8^{\circ}$ are 1.46, 1.89, 2.70, and 2.74 eV, respectively. The experimental $E_L$ is very close to the calculated $\Delta E_{vHs}$ corresponding to TBLG of twist angle $13.2^{\circ}$.  Therefore, our theoretical observation suggests that the closest resonance ($E_{L} \approx \Delta E_{vHs}$) for laser energy of 2.33 eV occurs at the vicinity of $\theta = 13.2^{\circ}$, consistent with our experimental result and $\theta_{C} = 12.3^{\circ}$ calculated based on continuum model.\cite{CVD_TBLG_JPCC} Hence, our theoretical study and experimental observation indicate that the resonance of $E_{L}$ with $\Delta E_{vHs}$ is the fundamental reason for the enhancement of the G peak in TBLG of a twist angle of $12.5^{\circ}$.

Apart from the first order Raman modes, we also investigated the second order peaks and combination modes which also yield crucial information regarding the layer stacking. Specifically, we have looked into the Raman spectrum in the range of 1700~cm$^{-1}$ to 2100~cm$^{-1}$ in the TBLG and compared with SLG and Bernal stacked BLG as shown in Fig.~4a. In BLG, a peak is observed at 1750~cm$^{-1}$ which is the M band, an overtone of oTO phonon giving rise to an electron double resonant intravalley scattering. This mode becomes Raman active due to interlayer coupling. Hence it is absent in SLG, but present in BLG, multilayer graphene and graphite with Bernal stacking.~\cite{Raman_2phononbands}  We find that the M-band is absent in TBLG samples clearly indicating deviations from Bernal stacking which renders the oTO phonon Raman inactive. The absence of M band indicates significant weakening of the interlayer interactions as compared to Bernal stacked BLG. Surprisingly, we observe the M-band for twist angle of $12.5^{\circ}$. It has been shown that M-band can selectively be seen in single-walled carbon nanotubes (SW-CNT) of certain chiralities and diameters when the laser excitation energy matches the transition energy between the valence and conduction band vHs.~\cite{CNT_1,CNT_2} On a similar note, the observation of the M-band coinciding with the G-peak resonance enhancement, suggests its origin related to the resonant excitation between vHs and inelastic scattering involving oTO phonons.

We also investigate the reported combination modes around 1860~cm$^{-1}$ and 2000~cm$^{-1}$ for the TBLG samples.~\cite{combination_modes,layer_stacking_combinationmodes} These modes are shown for TBLG samples along with SLG and BLG in Fig.~4a. The nomenclature of the various phonons giving rise to these combination modes are given in Table~1. At 1860~cm$^{-1}$, we observe that for the BLG, there are two convoluted peaks corresponding to LO$^{-}$TA (LOTA$@\Gamma$) and LO$^-$TO$^{\prime}$$@\Gamma$ which reduces to a single narrow peak attributed to  LOTA$@\Gamma$ in SLG. As the twist angle increases, we observe the reduction in intensity of the peak corresponding to LO$^-$TO$^{\prime}$$@\Gamma$. In BLG, around 2000~cm$^{-1}$, the phonon contributions around the $\Gamma$ and K points are comparable leading to a band which is a convolution of three peaks corresponding to scattering by the phonon modes LO$^+$LA$@\Gamma$, LO$^-$LO$^{\prime}$$@\Gamma$ and TOZO$@$K. The peak corresponding to TOZO$@$K vanishes for the SLG as a result of negligible electron-phonon coupling of ZO$^-$$@$K, while the interlayer interactions restore this mode in BLG.~\cite{Raman_2phononbands} At high twist angles, we observe the absence of the Raman peak from TOZO$@$K. Altogether, we find that above $12.5^{\circ}$ twist angle, the Raman modes in the range of 1700~cm$^{-1}$ to 2100~cm$^{-1}$ resemble that of SLG. This suggests the transition from a commensurate (C-TBLG) to an incommensurate (I-TBLG) sublattice with increasing rotational misorientation. We also observe a significant blue shift of the phonon modes associated with LOTA and LOLA  at higher twist angles when compared to that of SLG as shown in Fig.~4b. This shows a stiffening of the phonon modes possibly due to a compressive strain in I-TBLG as opposed to the C-TBLG. The blue shift of the combination modes also follow the same variation as that of 2D peak. Such blue shift of the 2D peak has also been observed earlier in h-BN encapsulated graphene with incommensurate superlattice structure.~\cite{graphene_hBN} $30^{\circ}$ TBLG is considered to be a quasicrystal with no long-range translational symmetry but rotational symmetry being preserved with a dodecagonal pattern of arrangement that can also be inferred from the reciprocal space as shown in Fig.~4d. The evolution from the 6-fold symmetry at $10^{\circ}$  to the 12-fold symmetry at $30^{\circ}$ is represented in the reciprocal space in Fig.~4d. From the combination modes Raman spectrum and symmetry analysis of the reciprocal space, we conclude that TBLG with twist angle of $20^{\circ}$ behaves as incommensurate lattice with no quasicrystalline symmetry. 

% TABLE 1 starts here 

 \begin{table}[h!]
 \centering
   \begin{tabular}{ | c | c | p{5cm} | }
    \hline
    Combination modes & Vibrational Raman modes & Significance \\ [0.5 ex]
    \hline 
      iLO + iTO or iLO$^+$ + iTO$^+$ & E$_g$ & Doubly degenerate in-plane symmetric with in-phase and out-of-phase displacement \\
    \hline
    ZO' and ZO$^+$ & A$_{1g}$ & Non-degenerate out-of-plane symmetric with in-phase and out-of-phase displacement \\
    \hline
    ZA and ZO$^-$ & A$_{2u}$ & Non-degenerate out-of-plane anti-symmetric with in-phase and out-of-phase displacement \\
    \hline
    LA + TA and LO$^-$ + TO$^-$ & E$_u$ & Doubly degenerate in-plane anti-symmetric with in-phase and out-of-phase displacement \\[1ex]
    \hline
    \end{tabular}
    \caption{Combination modes between 1860~cm$^{-1}$ and 2000~cm$^{-1}$ for the TBLG samples }
    \label {table:1}
    \end{table}

%table-1 ends here

Even though, the interlayer interactions are weakened in the I-TBLG, it is important to see if there is any contribution of the Moir\'e potential to the electron scattering mechanisms in the two layers. To study this we have looked for signatures of predicted R-peak arising from the inter-valley double resonant scattering. Here an excited electron is elastically scattering to another inequivalent valley by a large rotation momentum $q(\theta)$ and then inelastically backscattered to the same $k$ value by a zone boundary phonon $Q_{inter}$, followed by recombination with hole giving the Raman scattered photon.~\cite{Folded_Gr_4} Near the K-point, the TO phonon branch has very strong electron-phonon coupling and provide the momentum required for the inter-valley scattering. For a twist angles of $30^{\circ}$ and $20^{\circ}$, we observe low intensity peaks at $\sim$ 1375~cm$^{-1}$ and $\sim$1435~cm$^{-1}$ respectively in the Raman spectrum shown in Fig.~4c which also matches with the predicted values for the R peaks. This shows that even in incommensurate lattices, the Moir\'e potential can influence the electron scattering mechanisms.

\subsection{Conclusions}

In conclusion, we have carried out detailed Raman spectroscopic investigations of TBLG and find clear Raman signatures which can be correlated with the changes in the electronic bands due to the Moir\'e potential induced by twisting the graphene layers. When the laser excitation matches the energy separation between the vHs arising from hybridization between the bands of the two layers, resonance features arise in the G-peak and the M-band in TBLG. The Moir\'e potential can also induce intra- and inter-valley scattering processes giving rise to twist angle specific peaks in the Raman spectrum of TBLG.The observed upshift of the 2D peak and combination modes at high twist angles can be attributed to the compressive strain in incommensurate superlattice. The 2D peak intensity and FWHM, together with the Raman spectrum of the combination modes can jointly distinguish between commensurate and incommensurate superlattices formed in TBLG making this a valuable characterization technique to investigate the degree of commensurability in graphene-based 2D heterostructures.

\subsection{Methods}

%% experimental method and Raman equipment specification
\textbf{Raman measurements}

Raman scattering measurements were performed using the WITEC alpha300 system equipped with 1800 lines/mm grating. The excitation was done using a 532~nm diode laser with a power of approximately 16.87~mW and a spot size of 500~nm.
To observe the samples, a 100x objective lens with a 0.95 NA was utilized and the intensity was detected using a CCD. The data analysis was carried out using the WITEC Project software.

%% DFT data method specifications
\textbf{DFT calculations}

In order to gain insights into the changes in the interlayer interactions and electron-phonon scattering of TBLG with twist angle, the first-principles DFT calculations are performed using VASP.\cite{DFT7_ab_initio,DFT8_ab_initio} The all-electron projector augmented wave (PAW) potentials are used to represent the ion-electron interactions in the TBLG systems.\cite{DFT5_psuedopotential,DFT6_augmentedwave}  For the calculations, the electronic exchange and correlation part of the potential is described by the Perdew-Burke-Ernzerhof (PBE) generalized gradient approximation (GGA).\cite{DFT4_gradient_approach} A 20 Å vacuum is used along the c-axis in order to avoid spurious interactions between periodically repeated images of single layer and bilayer TBLG systems. The Kohn Sham orbitals are expanded using the plane wave basis sets with an energy cutoff of value 500 eV. All structures are relaxed using the conjugate-gradient algorithm until the Hellmann-Feynman forces on every atom are less than 0.005 eV${\angstrom}^{-1}$. For bilayer graphene and TBLG, the lattice parameters and atomic positions are optimized by considering the weak van der Waals (vdW) interactions between the layers as implemented in Grimme's PBE-D2 \cite{DFT3_GGA}. For relaxation, a well-converged Monkhorst-Pack (MP) k-grid of 12×12×1 was used to sample the Brillouin zone (BZ) of SLG and BLG.\cite{DFT2_special_points} For TBLG of twist angles $27.8^{\circ}$ and $21.8^{\circ}$, the k-grid of $8\times8\times1$ is used, and for $13.2^{\circ}$ and $9.43^{\circ}$ twist angles, a $6\times6\times1$ k-grid is used.

%% captions are here 

\subsection{FIGURES}   
    
\begin{figure}
\includegraphics[width=1\textwidth]{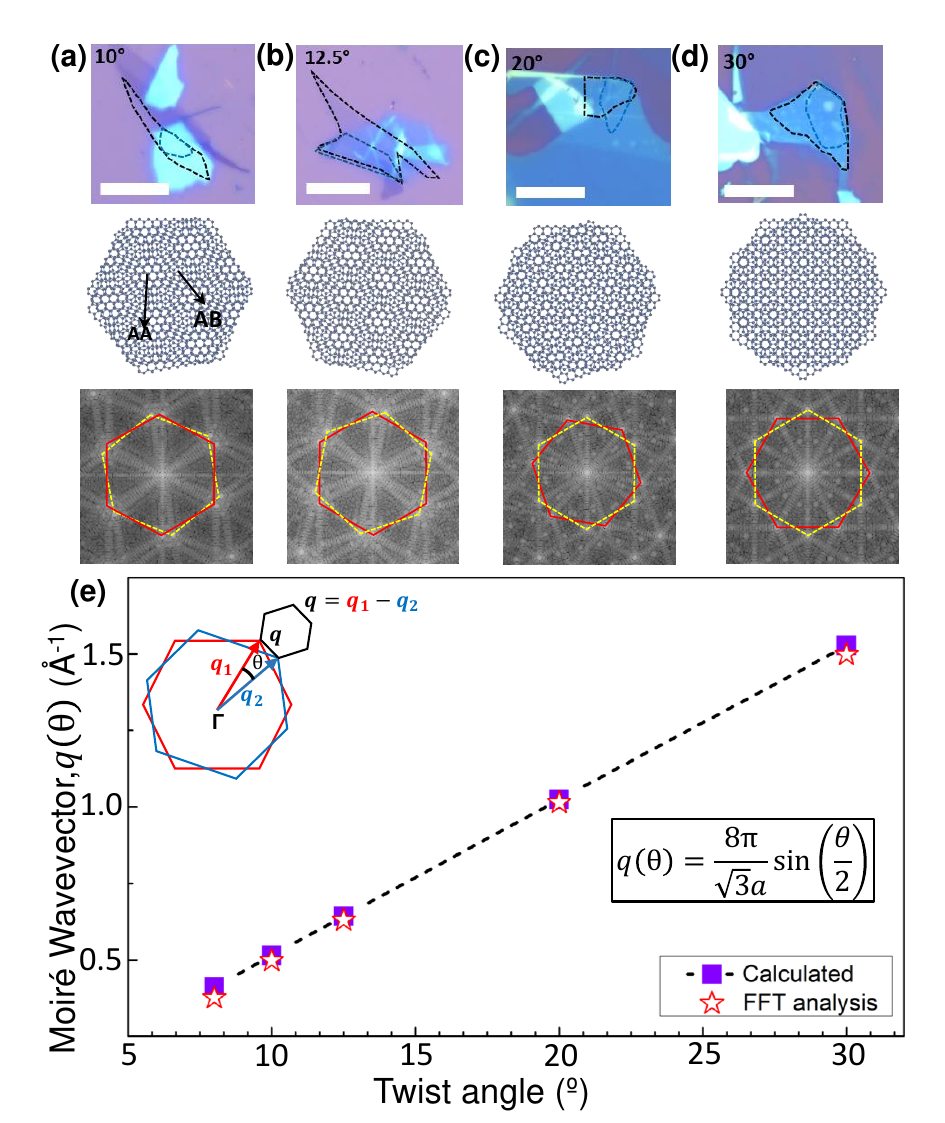}
\caption{(a) to (d) Optical images of samples (first row)(scale: 10 $\mu$m), real space lattice structures (second row) and FFT analysis with reciprocal space representing first Brillouin zone seperation (third row) for $10^{\circ}$, $12.5^{\circ}$, $20^{\circ}$ and $30^{\circ}$ TBLG samples. (e) Plot of $q(\theta)$ vs $\theta$ for twist angles ranging from $8^{\circ}$ to $30^{\circ}$ TBLG samples showing the values calculated by the formula in the inset and obtained from the FFT analysis of Moiré wave-vector $q(\theta)$ as shown in inset figure.}
\label{Figure 1 }
\end{figure}

\begin{figure}
\includegraphics[width=1\textwidth]{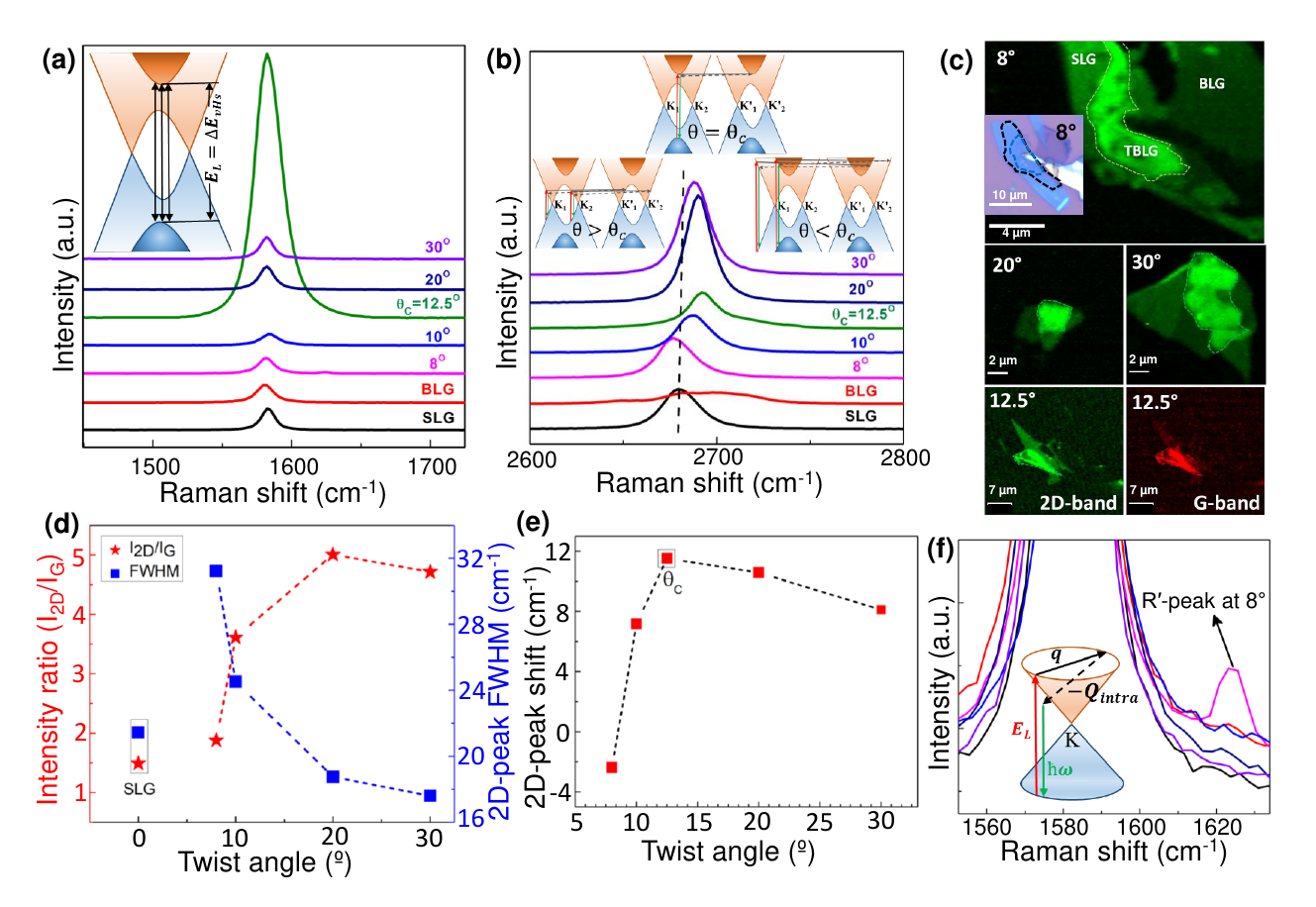}
\caption{(a) Raman shift of G-peak for all TBLG samples. G-peak enhancement occurs for $\theta_{C} = 12.5^{\circ}$ due to parallel band transitions (inset figure) at $E_{L} = \Delta E_{vHs}$ . (b) Raman shift of 2D-peak for all TBLG samples. The various intervalley scattering possibilities occuring for $\theta > \theta_{C}$, $\theta = \theta_{C}$ and $\theta < \theta_{C}$ are shown in the inset. (c) Raman maps of 2D peak (green) for different TBLG samples. Raman map for the G peak (red) shows a resonance enhancement for the $12.5^{\circ}$ TBLG. (d) Plot for $I_{2D} / I_G$  and 2D-peak FWHM as a function of twist angle, $\theta$ is shown for  different TBLG samples. (e) Plot of 2D-peak shift of TBLG relative to the SLG as a function of twist angle, $\theta$. (f) Appearance of R$^{\prime}$-peak for $8^{\circ}$ TBLG sample along with the intravalley scattering process due to Moir\'e potential as depicted in the inset.}
\label{Figure 2 }
\end{figure}
 
\begin{figure}
\includegraphics[width=1\textwidth]{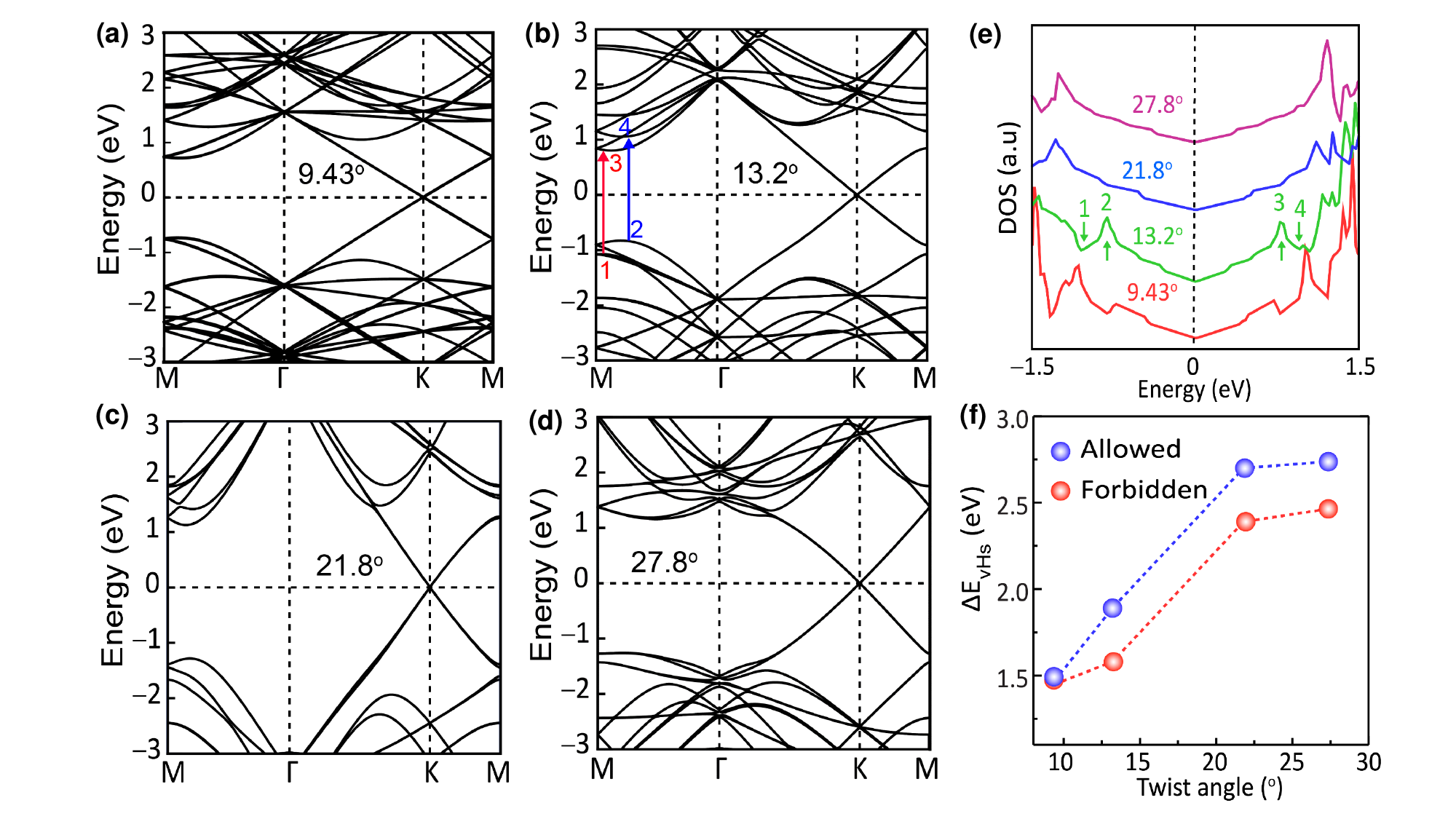}
\caption{Electronic band structures of TBLG for twist angles of (a) $9.43^{\circ}$, (b) $13.2^{\circ}$, (c) $21.8^{\circ}$ and (d) $27.8^{\circ}$, respectively. The red and blue arrows in Figure (b) represent the allowed (1→3 or 2→4) optical transitions. (e) The calculated total DOS of TBLG for twist angles $9.43^{\circ}$, $13.2^{\circ}$, $21.8^{\circ}$ and  $27.8^{\circ}$.The allowed (1→3 or 2→4) is given for $13.2^{\circ}$. (f) The value of $\Delta E_{vHs}$ of allowed (1→3 or 2→4) and lowest forbidden transitions (2→3) in TBLG for twist angles $9.43^{\circ}$, $13.2^{\circ}$, $21.8^{\circ}$ and $27.8^{\circ}$ respectively.}
\label{Figure 3 }
\end{figure} 

\begin{figure}
\includegraphics[width=1\textwidth]{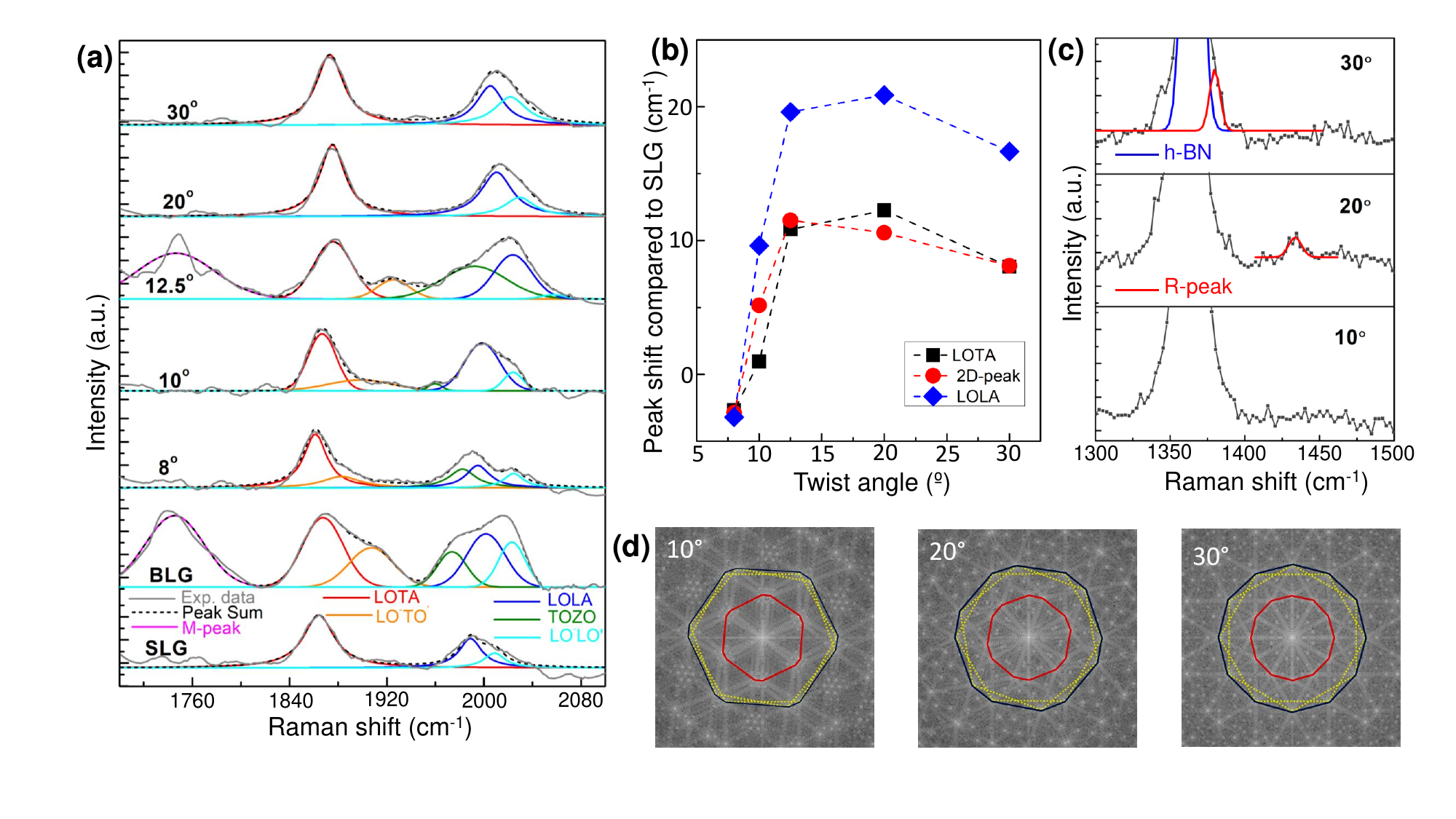}
\caption{(a) Plot of Raman shift for M-band and combination modes (near 1860~cm$^{-1}$ and 2000~cm$^{-1}$) in different TBLG samples. Peak-fitting of the data identifies all the possible combination modes. (b) Raman peak shift of TBLG compared to SLG for LOTA , LOLA and 2D peaks for different TBLG samples. (c) R-peak (in red) appears for $20^{\circ}$ and $30^{\circ}$ at 1435~cm$^{-1}$  and 1375~cm$^{-1}$ respectively. (d) In the Fourier space, the transition of TBLG system from commensurate lattice ($10^{\circ}$) to incommensurate lattice ($20^{\circ}$ and $30^{\circ}$) is shown. The rotational symmetry is near 6-fold for $10^{\circ}$ gradually reaching 12-fold for $30^{\circ}$ which is also a quasi-crystalline state.}
\label{Figure 4 }
\end{figure} 

\newpage

\begin{acknowledgement}

The authors acknowledge the funding from SERB Startup Research Grant and DST Nanomission, India. VK, NG and AR acknowledge funding from AOARD (Asian Office of Aerospace Research and Development) under Grant No. FA2386-21-1-4014.VK, VP, AMB and SP acknowledge the STEP facility at IIT Kharagpur.  SM, NM, and AKS thank the Materials Research Centre (MRC), Supercomputer Education and Research Centre (SERC), and Solid State and Structural Chemistry Unit (SSCU), Indian Institute of Science, Bangalore, for providing the required computational facilities. SM, NM, and AKS also acknowledge support from The Institute of Eminence (IoE) scheme of The
Ministry of Human Resource Development, Government of India. K.W. and T.T. acknowledge support from the JSPS KAKENHI (Grant Numbers 21H05233 and 23H02052) and World Premier International Research Center Initiative (WPI), MEXT, Japan.
\end{acknowledgement}

\bibliography{Manuscript}

\providecommand{\latin}[1]{#1}
\makeatletter
\providecommand{\doi}
  {\begingroup\let\do\@makeother\dospecials
  \catcode`\{=1 \catcode`\}=2 \doi@aux}
\providecommand{\doi@aux}[1]{\endgroup\texttt{#1}}
\makeatother
\providecommand*\mcitethebibliography{\thebibliography}
\csname @ifundefined\endcsname{endmcitethebibliography}
  {\let\endmcitethebibliography\endthebibliography}{}
\begin{mcitethebibliography}{70}
\providecommand*\natexlab[1]{#1}
\providecommand*\mciteSetBstSublistMode[1]{}
\providecommand*\mciteSetBstMaxWidthForm[2]{}
\providecommand*\mciteBstWouldAddEndPuncttrue
  {\def\EndOfBibitem{\unskip.}}
\providecommand*\mciteBstWouldAddEndPunctfalse
  {\let\EndOfBibitem\relax}
\providecommand*\mciteSetBstMidEndSepPunct[3]{}
\providecommand*\mciteSetBstSublistLabelBeginEnd[3]{}
\providecommand*\EndOfBibitem{}
\mciteSetBstSublistMode{f}
\mciteSetBstMaxWidthForm{subitem}{(\alph{mcitesubitemcount})}
\mciteSetBstSublistLabelBeginEnd
  {\mcitemaxwidthsubitemform\space}
  {\relax}
  {\relax}

\bibitem[Cao \latin{et~al.}(2018)Cao, Fatemi, Fang, Watanabe, Taniguchi,
  Kaxiras, and Jarillo-Herrero]{MATBLG_superconductivity}
Cao,~Y.; Fatemi,~V.; Fang,~S.; Watanabe,~K.; Taniguchi,~T.; Kaxiras,~E.;
  Jarillo-Herrero,~P. Unconventional superconductivity in magic-angle graphene
  superlattices. \emph{Nature} \textbf{2018}, \emph{556}, 43--50\relax
\mciteBstWouldAddEndPuncttrue
\mciteSetBstMidEndSepPunct{\mcitedefaultmidpunct}
{\mcitedefaultendpunct}{\mcitedefaultseppunct}\relax
\EndOfBibitem
\bibitem[Cao \latin{et~al.}(2018)Cao, Fatemi, Demir, Fang, Tomarken, Luo,
  Sanchez-Yamagishi, Watanabe, Taniguchi, Kaxiras, Ashoori, and
  Jarillo-Herrero]{MATBLG_correlated_insulator}
Cao,~Y.; Fatemi,~V.; Demir,~A.; Fang,~S.; Tomarken,~S.~L.; Luo,~J.~Y.;
  Sanchez-Yamagishi,~J.~D.; Watanabe,~K.; Taniguchi,~T.; Kaxiras,~E.;
  Ashoori,~R.~C.; Jarillo-Herrero,~P. Correlated insulator behaviour at
  half-filling in magic-angle graphene superlattices. \emph{Nature}
  \textbf{2018}, \emph{556}, 80--84\relax
\mciteBstWouldAddEndPuncttrue
\mciteSetBstMidEndSepPunct{\mcitedefaultmidpunct}
{\mcitedefaultendpunct}{\mcitedefaultseppunct}\relax
\EndOfBibitem
\bibitem[Sharpe \latin{et~al.}(2019)Sharpe, Fox, Barnard, Finney, Watanabe,
  Taniguchi, Kastner, and Goldhaber-Gordon]{TBLG_ferromagnetism}
Sharpe,~A.~L.; Fox,~E.~J.; Barnard,~A.~W.; Finney,~J.; Watanabe,~K.;
  Taniguchi,~T.; Kastner,~M.~A.; Goldhaber-Gordon,~D. Emergent ferromagnetism
  near three-quarters filling in twisted bilayer graphene. \emph{Science}
  \textbf{2019}, \emph{365}, 605--608\relax
\mciteBstWouldAddEndPuncttrue
\mciteSetBstMidEndSepPunct{\mcitedefaultmidpunct}
{\mcitedefaultendpunct}{\mcitedefaultseppunct}\relax
\EndOfBibitem
\bibitem[Serlin \latin{et~al.}(2020)Serlin, Tschirhart, Polshyn, Zhang, Zhu,
  Watanabe, Taniguchi, Balents, and Young]{TBLG_AHE}
Serlin,~M.; Tschirhart,~C.~L.; Polshyn,~H.; Zhang,~Y.; Zhu,~J.; Watanabe,~K.;
  Taniguchi,~T.; Balents,~L.; Young,~A.~F. Intrinsic quantized anomalous Hall
  effect in a moiré heterostructure. \emph{Science} \textbf{2020}, \emph{367},
  900--903\relax
\mciteBstWouldAddEndPuncttrue
\mciteSetBstMidEndSepPunct{\mcitedefaultmidpunct}
{\mcitedefaultendpunct}{\mcitedefaultseppunct}\relax
\EndOfBibitem
\bibitem[Bistritzer and MacDonald(2011)Bistritzer, and
  MacDonald]{Moirebands_PNAS}
Bistritzer,~R.; MacDonald,~A.~H. Moiré bands in twisted double-layer graphene.
  \emph{Proceedings of the National Academy of Sciences} \textbf{2011},
  \emph{108}, 12233--12237\relax
\mciteBstWouldAddEndPuncttrue
\mciteSetBstMidEndSepPunct{\mcitedefaultmidpunct}
{\mcitedefaultendpunct}{\mcitedefaultseppunct}\relax
\EndOfBibitem
\bibitem[Yoo \latin{et~al.}(2019)Yoo, Engelke, Carr, Fang, Zhang, Cazeaux,
  Sung, Hovden, Tsen, Taniguchi, Watanabe, Yi, Kim, Luskin, Tadmor, Kaxiras,
  and Kim]{vdW_elastic_energy_TBLG}
Yoo,~H. \latin{et~al.}  Atomic and electronic reconstruction at the van der
  Waals interface in twisted bilayer graphene. \emph{Nature Materials}
  \textbf{2019}, \emph{18}, 448--453\relax
\mciteBstWouldAddEndPuncttrue
\mciteSetBstMidEndSepPunct{\mcitedefaultmidpunct}
{\mcitedefaultendpunct}{\mcitedefaultseppunct}\relax
\EndOfBibitem
\bibitem[Ahn \latin{et~al.}(2018)Ahn, Moon, Kim, Kim, Shin, Kim, Cha, Kahng,
  Kim, Koshino, Son, Yang, and Ahn]{QCgraphene_Science}
Ahn,~S.~J.; Moon,~P.; Kim,~T.-H.; Kim,~H.-W.; Shin,~H.-C.; Kim,~E.~H.;
  Cha,~H.~W.; Kahng,~S.-J.; Kim,~P.; Koshino,~M.; Son,~Y.-W.; Yang,~C.-W.;
  Ahn,~J.~R. Dirac electrons in a dodecagonal graphene quasicrystal.
  \emph{Science} \textbf{2018}, \emph{361}, 782--786\relax
\mciteBstWouldAddEndPuncttrue
\mciteSetBstMidEndSepPunct{\mcitedefaultmidpunct}
{\mcitedefaultendpunct}{\mcitedefaultseppunct}\relax
\EndOfBibitem
\bibitem[Yao \latin{et~al.}(2018)Yao, Wang, Bao, Zhang, Zhang, Bao, Chan, Chen,
  Avila, Asensio, Zhu, and Zhou]{QCgraphene_PNAS}
Yao,~W.; Wang,~E.; Bao,~C.; Zhang,~Y.; Zhang,~K.; Bao,~K.; Chan,~C.~K.;
  Chen,~C.; Avila,~J.; Asensio,~M.~C.; Zhu,~J.; Zhou,~S. Quasicrystalline 30°
  twisted bilayer graphene as an incommensurate superlattice with strong
  interlayer coupling. \emph{Proceedings of the National Academy of Sciences}
  \textbf{2018}, \emph{115}, 6928--6933\relax
\mciteBstWouldAddEndPuncttrue
\mciteSetBstMidEndSepPunct{\mcitedefaultmidpunct}
{\mcitedefaultendpunct}{\mcitedefaultseppunct}\relax
\EndOfBibitem
\bibitem[Pezzini \latin{et~al.}(2020)Pezzini, Mišeikis, Piccinini, Forti,
  Pace, Engelke, Rossella, Watanabe, Taniguchi, Kim, and
  Coletti]{QCgraphene_Nanoletters}
Pezzini,~S.; Mišeikis,~V.; Piccinini,~G.; Forti,~S.; Pace,~S.; Engelke,~R.;
  Rossella,~F.; Watanabe,~K.; Taniguchi,~T.; Kim,~P.; Coletti,~C. 30°-Twisted
  Bilayer Graphene Quasicrystals from Chemical Vapor Deposition. \emph{Nano
  Letters} \textbf{2020}, \emph{20}, 3313--3319\relax
\mciteBstWouldAddEndPuncttrue
\mciteSetBstMidEndSepPunct{\mcitedefaultmidpunct}
{\mcitedefaultendpunct}{\mcitedefaultseppunct}\relax
\EndOfBibitem
\bibitem[Utama \latin{et~al.}(2021)Utama, Koch, Lee, Leconte, Li, Zhao, Jiang,
  Zhu, Watanabe, Taniguchi, Ashby, Weber-Bargioni, Zettl, Jozwiak, Jung,
  Rotenberg, Bostwick, and Wang]{STM_1}
Utama,~M. I.~B. \latin{et~al.}  Visualization of the flat electronic band in
  twisted bilayer graphene near the magic angle twist. \emph{Nature Physics}
  \textbf{2021}, \emph{17}, 184--188\relax
\mciteBstWouldAddEndPuncttrue
\mciteSetBstMidEndSepPunct{\mcitedefaultmidpunct}
{\mcitedefaultendpunct}{\mcitedefaultseppunct}\relax
\EndOfBibitem
\bibitem[Jiang \latin{et~al.}(2019)Jiang, Lai, Watanabe, Taniguchi, Haule, Mao,
  and Andrei]{STM_2}
Jiang,~Y.; Lai,~X.; Watanabe,~K.; Taniguchi,~T.; Haule,~K.; Mao,~J.;
  Andrei,~E.~Y. Charge order and broken rotational symmetry in magic-angle
  twisted bilayer graphene. \emph{Nature} \textbf{2019}, \emph{573},
  91--95\relax
\mciteBstWouldAddEndPuncttrue
\mciteSetBstMidEndSepPunct{\mcitedefaultmidpunct}
{\mcitedefaultendpunct}{\mcitedefaultseppunct}\relax
\EndOfBibitem
\bibitem[Kerelsky \latin{et~al.}(2019)Kerelsky, McGilly, Kennes, Xian,
  Yankowitz, Chen, Watanabe, Taniguchi, Hone, Dean, Rubio, and
  Pasupathy]{STM_3}
Kerelsky,~A.; McGilly,~L.~J.; Kennes,~D.~M.; Xian,~L.; Yankowitz,~M.; Chen,~S.;
  Watanabe,~K.; Taniguchi,~T.; Hone,~J.; Dean,~C.; Rubio,~A.; Pasupathy,~A.~N.
  Maximized electron interactions at the magic angle in twisted bilayer
  graphene. \emph{Nature} \textbf{2019}, \emph{572}, 95--100\relax
\mciteBstWouldAddEndPuncttrue
\mciteSetBstMidEndSepPunct{\mcitedefaultmidpunct}
{\mcitedefaultendpunct}{\mcitedefaultseppunct}\relax
\EndOfBibitem
\bibitem[Choi \latin{et~al.}(2019)Choi, Kemmer, Peng, Thomson, Arora, Polski,
  Zhang, Ren, Alicea, Refael, von Oppen, Watanabe, Taniguchi, and
  Nadj-Perge]{STM_4}
Choi,~Y.; Kemmer,~J.; Peng,~Y.; Thomson,~A.; Arora,~H.; Polski,~R.; Zhang,~Y.;
  Ren,~H.; Alicea,~J.; Refael,~G.; von Oppen,~F.; Watanabe,~K.; Taniguchi,~T.;
  Nadj-Perge,~S. Electronic correlations in twisted bilayer graphene near the
  magic angle. \emph{Nature Physics} \textbf{2019}, \emph{15}, 1174--1180\relax
\mciteBstWouldAddEndPuncttrue
\mciteSetBstMidEndSepPunct{\mcitedefaultmidpunct}
{\mcitedefaultendpunct}{\mcitedefaultseppunct}\relax
\EndOfBibitem
\bibitem[Xie \latin{et~al.}(2019)Xie, Lian, Jäck, Liu, Chiu, Watanabe,
  Taniguchi, Bernevig, and Yazdani]{STM_5}
Xie,~Y.; Lian,~B.; Jäck,~B.; Liu,~X.; Chiu,~C.-L.; Watanabe,~K.;
  Taniguchi,~T.; Bernevig,~B.~A.; Yazdani,~A. Spectroscopic signatures of
  many-body correlations in magic-angle twisted bilayer graphene. \emph{Nature}
  \textbf{2019}, \emph{572}, 101--105\relax
\mciteBstWouldAddEndPuncttrue
\mciteSetBstMidEndSepPunct{\mcitedefaultmidpunct}
{\mcitedefaultendpunct}{\mcitedefaultseppunct}\relax
\EndOfBibitem
\bibitem[Wong \latin{et~al.}(2020)Wong, Nuckolls, Oh, Lian, Xie, Jeon,
  Watanabe, Taniguchi, Bernevig, and Yazdani]{STM_6}
Wong,~D.; Nuckolls,~K.~P.; Oh,~M.; Lian,~B.; Xie,~Y.; Jeon,~S.; Watanabe,~K.;
  Taniguchi,~T.; Bernevig,~B.~A.; Yazdani,~A. Cascade of electronic transitions
  in magic-angle twisted bilayer graphene. \emph{Nature} \textbf{2020},
  \emph{582}, 198--202\relax
\mciteBstWouldAddEndPuncttrue
\mciteSetBstMidEndSepPunct{\mcitedefaultmidpunct}
{\mcitedefaultendpunct}{\mcitedefaultseppunct}\relax
\EndOfBibitem
\bibitem[Zondiner \latin{et~al.}(2020)Zondiner, Rozen, Rodan-Legrain, Cao,
  Queiroz, Taniguchi, Watanabe, Oreg, von Oppen, Stern, Berg, Jarillo-Herrero,
  and Ilani]{STM_7}
Zondiner,~U.; Rozen,~A.; Rodan-Legrain,~D.; Cao,~Y.; Queiroz,~R.;
  Taniguchi,~T.; Watanabe,~K.; Oreg,~Y.; von Oppen,~F.; Stern,~A.; Berg,~E.;
  Jarillo-Herrero,~P.; Ilani,~S. Cascade of phase transitions and Dirac
  revivals in magic-angle graphene. \emph{Nature} \textbf{2020}, \emph{582},
  203--208\relax
\mciteBstWouldAddEndPuncttrue
\mciteSetBstMidEndSepPunct{\mcitedefaultmidpunct}
{\mcitedefaultendpunct}{\mcitedefaultseppunct}\relax
\EndOfBibitem
\bibitem[Choi \latin{et~al.}(2021)Choi, Kim, Peng, Thomson, Lewandowski,
  Polski, Zhang, Arora, Watanabe, Taniguchi, Alicea, and Nadj-Perge]{STM_8}
Choi,~Y.; Kim,~H.; Peng,~Y.; Thomson,~A.; Lewandowski,~C.; Polski,~R.;
  Zhang,~Y.; Arora,~H.~S.; Watanabe,~K.; Taniguchi,~T.; Alicea,~J.;
  Nadj-Perge,~S. Correlation-driven topological phases in magic-angle twisted
  bilayer graphene. \emph{Nature} \textbf{2021}, \emph{589}, 536--541\relax
\mciteBstWouldAddEndPuncttrue
\mciteSetBstMidEndSepPunct{\mcitedefaultmidpunct}
{\mcitedefaultendpunct}{\mcitedefaultseppunct}\relax
\EndOfBibitem
\bibitem[Nuckolls \latin{et~al.}(2020)Nuckolls, Oh, Wong, Lian, Watanabe,
  Taniguchi, Bernevig, and Yazdani]{STM_9}
Nuckolls,~K.~P.; Oh,~M.; Wong,~D.; Lian,~B.; Watanabe,~K.; Taniguchi,~T.;
  Bernevig,~B.~A.; Yazdani,~A. Strongly correlated Chern insulators in
  magic-angle twisted bilayer graphene. \emph{Nature} \textbf{2020},
  \emph{588}, 610--615\relax
\mciteBstWouldAddEndPuncttrue
\mciteSetBstMidEndSepPunct{\mcitedefaultmidpunct}
{\mcitedefaultendpunct}{\mcitedefaultseppunct}\relax
\EndOfBibitem
\bibitem[Lisi \latin{et~al.}(2021)Lisi, Lu, Benschop, de~Jong, Stepanov, Duran,
  Margot, Cucchi, Cappelli, Hunter, Tamai, Kandyba, Giampietri, Barinov, Jobst,
  Stalman, Leeuwenhoek, Watanabe, Taniguchi, Rademaker, van~der Molen, Allan,
  Efetov, and Baumberger]{ARPES_1}
Lisi,~S. \latin{et~al.}  Observation of flat bands in twisted bilayer graphene.
  \emph{Nature Physics} \textbf{2021}, \emph{17}, 189--193\relax
\mciteBstWouldAddEndPuncttrue
\mciteSetBstMidEndSepPunct{\mcitedefaultmidpunct}
{\mcitedefaultendpunct}{\mcitedefaultseppunct}\relax
\EndOfBibitem
\bibitem[Sato \latin{et~al.}(2021)Sato, Hayashi, Ito, Masago, Takamura,
  Morimoto, Maekawa, Lee, Qiao, Kim, Nakagahara, Wakabayashi, Hibino, and
  Norimatsu]{ARPES_2}
Sato,~K.; Hayashi,~N.; Ito,~T.; Masago,~N.; Takamura,~M.; Morimoto,~M.;
  Maekawa,~T.; Lee,~D.; Qiao,~K.; Kim,~J.; Nakagahara,~K.; Wakabayashi,~K.;
  Hibino,~H.; Norimatsu,~W. Observation of a flat band and bandgap in
  millimeter-scale twisted bilayer graphene. \emph{Communications Materials}
  \textbf{2021}, \emph{2}, 117\relax
\mciteBstWouldAddEndPuncttrue
\mciteSetBstMidEndSepPunct{\mcitedefaultmidpunct}
{\mcitedefaultendpunct}{\mcitedefaultseppunct}\relax
\EndOfBibitem
\bibitem[Ferrari \latin{et~al.}(2006)Ferrari, Meyer, Scardaci, Casiraghi,
  Lazzeri, Mauri, Piscanec, Jiang, Novoselov, Roth, and Geim]{Ferrari_Raman}
Ferrari,~A.~C.; Meyer,~J.~C.; Scardaci,~V.; Casiraghi,~C.; Lazzeri,~M.;
  Mauri,~F.; Piscanec,~S.; Jiang,~D.; Novoselov,~K.~S.; Roth,~S.; Geim,~A.~K.
  Raman Spectrum of Graphene and Graphene Layers. \emph{Physical Review
  Letters} \textbf{2006}, \emph{97}, 187401\relax
\mciteBstWouldAddEndPuncttrue
\mciteSetBstMidEndSepPunct{\mcitedefaultmidpunct}
{\mcitedefaultendpunct}{\mcitedefaultseppunct}\relax
\EndOfBibitem
\bibitem[Ferrari(2007)]{Raman_disorder_1}
Ferrari,~A.~C. Raman spectroscopy of graphene and graphite: Disorder,
  electron–phonon coupling, doping and nonadiabatic effects. \emph{Solid
  State Communications} \textbf{2007}, \emph{143}, 47--57\relax
\mciteBstWouldAddEndPuncttrue
\mciteSetBstMidEndSepPunct{\mcitedefaultmidpunct}
{\mcitedefaultendpunct}{\mcitedefaultseppunct}\relax
\EndOfBibitem
\bibitem[Eckmann \latin{et~al.}(2012)Eckmann, Felten, Mishchenko, Britnell,
  Krupke, Novoselov, and Casiraghi]{Raman_disorder_2}
Eckmann,~A.; Felten,~A.; Mishchenko,~A.; Britnell,~L.; Krupke,~R.;
  Novoselov,~K.~S.; Casiraghi,~C. Probing the Nature of Defects in Graphene by
  Raman Spectroscopy. \emph{Nano Letters} \textbf{2012}, \emph{12},
  3925--3930\relax
\mciteBstWouldAddEndPuncttrue
\mciteSetBstMidEndSepPunct{\mcitedefaultmidpunct}
{\mcitedefaultendpunct}{\mcitedefaultseppunct}\relax
\EndOfBibitem
\bibitem[Lui \latin{et~al.}(2011)Lui, Li, Chen, Klimov, Brus, and
  Heinz]{Raman_stacking}
Lui,~C.~H.; Li,~Z.; Chen,~Z.; Klimov,~P.~V.; Brus,~L.~E.; Heinz,~T.~F. Imaging
  Stacking Order in Few-Layer Graphene. \emph{Nano Letters} \textbf{2011},
  \emph{11}, 164--169\relax
\mciteBstWouldAddEndPuncttrue
\mciteSetBstMidEndSepPunct{\mcitedefaultmidpunct}
{\mcitedefaultendpunct}{\mcitedefaultseppunct}\relax
\EndOfBibitem
\bibitem[Poncharal \latin{et~al.}(2008)Poncharal, Ayari, Michel, and
  Sauvajol]{Raman_misorientedBLG}
Poncharal,~P.; Ayari,~A.; Michel,~T.; Sauvajol,~J.-L. Raman spectra of
  misoriented bilayer graphene. \emph{Physical Review B} \textbf{2008},
  \emph{78}, 113407\relax
\mciteBstWouldAddEndPuncttrue
\mciteSetBstMidEndSepPunct{\mcitedefaultmidpunct}
{\mcitedefaultendpunct}{\mcitedefaultseppunct}\relax
\EndOfBibitem
\bibitem[Cong \latin{et~al.}(2011)Cong, Yu, Sato, Shang, Saito, Dresselhaus,
  and Dresselhaus]{Raman_TLG}
Cong,~C.; Yu,~T.; Sato,~K.; Shang,~J.; Saito,~R.; Dresselhaus,~G.~F.;
  Dresselhaus,~M.~S. Raman Characterization of ABA- and ABC-Stacked Trilayer
  Graphene. \emph{ACS Nano} \textbf{2011}, \emph{5}, 8760--8768\relax
\mciteBstWouldAddEndPuncttrue
\mciteSetBstMidEndSepPunct{\mcitedefaultmidpunct}
{\mcitedefaultendpunct}{\mcitedefaultseppunct}\relax
\EndOfBibitem
\bibitem[Podila \latin{et~al.}(2012)Podila, Rao, Tsuchikawa, Ishigami, and
  Rao]{Folded_Gr_1}
Podila,~R.; Rao,~R.; Tsuchikawa,~R.; Ishigami,~M.; Rao,~A.~M. Raman
  Spectroscopy of Folded and Scrolled Graphene. \emph{ACS Nano} \textbf{2012},
  \emph{6}, 5784--5790\relax
\mciteBstWouldAddEndPuncttrue
\mciteSetBstMidEndSepPunct{\mcitedefaultmidpunct}
{\mcitedefaultendpunct}{\mcitedefaultseppunct}\relax
\EndOfBibitem
\bibitem[Gupta \latin{et~al.}(2010)Gupta, Tang, Crespi, and
  Eklund]{Folded_Gr_3}
Gupta,~A.~K.; Tang,~Y.; Crespi,~V.~H.; Eklund,~P.~C. Nondispersive Raman D band
  activated by well-ordered interlayer interactions in rotationally stacked
  bilayer graphene. \emph{Physical Review B} \textbf{2010}, \emph{82},
  241406\relax
\mciteBstWouldAddEndPuncttrue
\mciteSetBstMidEndSepPunct{\mcitedefaultmidpunct}
{\mcitedefaultendpunct}{\mcitedefaultseppunct}\relax
\EndOfBibitem
\bibitem[Carozo \latin{et~al.}(2011)Carozo, Almeida, Ferreira, Cançado,
  Achete, and Jorio]{Folded_Gr_4}
Carozo,~V.; Almeida,~C.~M.; Ferreira,~E. H.~M.; Cançado,~L.~G.; Achete,~C.~A.;
  Jorio,~A. Raman Signature of Graphene Superlattices. \emph{Nano Letters}
  \textbf{2011}, \emph{11}, 4527--4534\relax
\mciteBstWouldAddEndPuncttrue
\mciteSetBstMidEndSepPunct{\mcitedefaultmidpunct}
{\mcitedefaultendpunct}{\mcitedefaultseppunct}\relax
\EndOfBibitem
\bibitem[Ramnani \latin{et~al.}(2017)Ramnani, Neupane, Ge, Balandin, Lake, and
  Mulchandani]{CVD_TBLG_carbon}
Ramnani,~P.; Neupane,~M.~R.; Ge,~S.; Balandin,~A.~A.; Lake,~R.~K.;
  Mulchandani,~A. Raman spectra of twisted CVD bilayer graphene. \emph{Carbon}
  \textbf{2017}, \emph{123}, 302--306\relax
\mciteBstWouldAddEndPuncttrue
\mciteSetBstMidEndSepPunct{\mcitedefaultmidpunct}
{\mcitedefaultendpunct}{\mcitedefaultseppunct}\relax
\EndOfBibitem
\bibitem[Lu \latin{et~al.}(2013)Lu, Lin, Liu, Yeh, Suenaga, and
  Chiu]{CVD_TBLG_ACSNano}
Lu,~C.-C.; Lin,~Y.-C.; Liu,~Z.; Yeh,~C.-H.; Suenaga,~K.; Chiu,~P.-W. Twisting
  Bilayer Graphene Superlattices. \emph{ACS Nano} \textbf{2013}, \emph{7},
  2587--2594\relax
\mciteBstWouldAddEndPuncttrue
\mciteSetBstMidEndSepPunct{\mcitedefaultmidpunct}
{\mcitedefaultendpunct}{\mcitedefaultseppunct}\relax
\EndOfBibitem
\bibitem[Zhou \latin{et~al.}(2021)Zhou, Leng, Liu, Yang, Liu, Liu, Zhao, Liu,
  Wang, Shang, Li, Zhao, Liu, and Xu]{CVD_TBLG_nanophotonics}
Zhou,~W.-G.; Leng,~Y.-C.; Liu,~L.-X.; Yang,~M.-M.; Liu,~W.; Liu,~J.-L.;
  Zhao,~P.; Liu,~Y.; Wang,~L.-L.; Shang,~Y.-X.; Li,~X.-L.; Zhao,~X.-H.;
  Liu,~X.-L.; Xu,~Y. Twist angle dependent absorption feature induced by
  interlayer rotations in CVD bilayer graphene. \emph{Nanophotonics}
  \textbf{2021}, \emph{10}, 2695--2703\relax
\mciteBstWouldAddEndPuncttrue
\mciteSetBstMidEndSepPunct{\mcitedefaultmidpunct}
{\mcitedefaultendpunct}{\mcitedefaultseppunct}\relax
\EndOfBibitem
\bibitem[He \latin{et~al.}(2013)He, Chung, Delaney, Keiser, Jauregui, Shand,
  Chancey, Wang, Bao, and Chen]{CVD_TBLG_lowfreqmodes}
He,~R.; Chung,~T.-F.; Delaney,~C.; Keiser,~C.; Jauregui,~L.~A.; Shand,~P.~M.;
  Chancey,~C.~C.; Wang,~Y.; Bao,~J.; Chen,~Y.~P. Observation of Low Energy
  Raman Modes in Twisted Bilayer Graphene. \emph{Nano Letters} \textbf{2013},
  \emph{13}, 3594--3601\relax
\mciteBstWouldAddEndPuncttrue
\mciteSetBstMidEndSepPunct{\mcitedefaultmidpunct}
{\mcitedefaultendpunct}{\mcitedefaultseppunct}\relax
\EndOfBibitem
\bibitem[Moutinho \latin{et~al.}(2021)Moutinho, Venezuela, and
  Pimenta]{CVD_TBLG_SSC}
Moutinho,~M. V.~O.; Venezuela,~P.; Pimenta,~M.~A. Raman Spectroscopy of Twisted
  Bilayer Graphene. \emph{C} \textbf{2021}, \emph{7}, 10\relax
\mciteBstWouldAddEndPuncttrue
\mciteSetBstMidEndSepPunct{\mcitedefaultmidpunct}
{\mcitedefaultendpunct}{\mcitedefaultseppunct}\relax
\EndOfBibitem
\bibitem[Xu \latin{et~al.}(2022)Xu, Hao, Huang, Zhao, Yang, Zhang, and
  Tong]{CVD_TBLG_JPCC}
Xu,~B.; Hao,~H.; Huang,~J.; Zhao,~Y.; Yang,~T.; Zhang,~J.; Tong,~L.
  Twist-Induced New Phonon Scattering Pathways in Bilayer Graphene Probed by
  Helicity-Resolved Raman Spectroscopy. \emph{The Journal of Physical Chemistry
  C} \textbf{2022}, \emph{126}, 10487--10493\relax
\mciteBstWouldAddEndPuncttrue
\mciteSetBstMidEndSepPunct{\mcitedefaultmidpunct}
{\mcitedefaultendpunct}{\mcitedefaultseppunct}\relax
\EndOfBibitem
\bibitem[Yeh \latin{et~al.}(2014)Yeh, Lin, Nayak, Lu, Liu, Suenaga, and
  Chiu]{CVD_TBLG_JRamanSpec}
Yeh,~C.-H.; Lin,~Y.-C.; Nayak,~P.~K.; Lu,~C.-C.; Liu,~Z.; Suenaga,~K.;
  Chiu,~P.-W. Probing interlayer coupling in twisted single-crystal bilayer
  graphene by Raman spectroscopy. \emph{Journal of Raman Spectroscopy}
  \textbf{2014}, \emph{45}, 912--917\relax
\mciteBstWouldAddEndPuncttrue
\mciteSetBstMidEndSepPunct{\mcitedefaultmidpunct}
{\mcitedefaultendpunct}{\mcitedefaultseppunct}\relax
\EndOfBibitem
\bibitem[Cong and Yu(2014)Cong, and Yu]{Folded_Gr_2}
Cong,~C.; Yu,~T. Evolution of Raman G and G' (2D) modes in folded graphene
  layers. \emph{Physical Review B} \textbf{2014}, \emph{89}, 235430\relax
\mciteBstWouldAddEndPuncttrue
\mciteSetBstMidEndSepPunct{\mcitedefaultmidpunct}
{\mcitedefaultendpunct}{\mcitedefaultseppunct}\relax
\EndOfBibitem
\bibitem[Havener \latin{et~al.}(2012)Havener, Zhuang, Brown, Hennig, and
  Park]{CVD_TBLG_angleresolved_Raman}
Havener,~R.~W.; Zhuang,~H.; Brown,~L.; Hennig,~R.~G.; Park,~J. Angle-Resolved
  Raman Imaging of Interlayer Rotations and Interactions in Twisted Bilayer
  Graphene. \emph{Nano Letters} \textbf{2012}, \emph{12}, 3162--3167\relax
\mciteBstWouldAddEndPuncttrue
\mciteSetBstMidEndSepPunct{\mcitedefaultmidpunct}
{\mcitedefaultendpunct}{\mcitedefaultseppunct}\relax
\EndOfBibitem
\bibitem[Kim \latin{et~al.}(2012)Kim, Coh, Tan, Regan, Yuk, Chatterjee,
  Crommie, Cohen, Louie, and Zettl]{AS_TBLG_PRL}
Kim,~K.; Coh,~S.; Tan,~L.~Z.; Regan,~W.; Yuk,~J.~M.; Chatterjee,~E.;
  Crommie,~M.~F.; Cohen,~M.~L.; Louie,~S.~G.; Zettl,~A. Raman Spectroscopy
  Study of Rotated Double-Layer Graphene: Misorientation-Angle Dependence of
  Electronic Structure. \emph{Physical Review Letters} \textbf{2012},
  \emph{108}, 246103\relax
\mciteBstWouldAddEndPuncttrue
\mciteSetBstMidEndSepPunct{\mcitedefaultmidpunct}
{\mcitedefaultendpunct}{\mcitedefaultseppunct}\relax
\EndOfBibitem
\bibitem[Ni \latin{et~al.}(2008)Ni, Wang, Yu, You, and Shen]{Folded_Gr_red_vF}
Ni,~Z.; Wang,~Y.; Yu,~T.; You,~Y.; Shen,~Z. Reduction of Fermi velocity in
  folded graphene observed by resonance Raman spectroscopy. \emph{Physical
  Review B} \textbf{2008}, \emph{77}, 235403\relax
\mciteBstWouldAddEndPuncttrue
\mciteSetBstMidEndSepPunct{\mcitedefaultmidpunct}
{\mcitedefaultendpunct}{\mcitedefaultseppunct}\relax
\EndOfBibitem
\bibitem[Wang \latin{et~al.}(2013)Wang, Su, Wu, Nie, Xie, Gong, Guo, Lee, Xing,
  Lu, Wang, Lu, McCarty, shem Pei, Robles-Hernandez, Hadjiev, and
  Bao]{CVD_TBLG_APL}
Wang,~Y. \latin{et~al.}  Resonance Raman spectroscopy of G-line and folded
  phonons in twisted bilayer graphene with large rotation angles. \emph{Applied
  Physics Letters} \textbf{2013}, \emph{103}, 123101\relax
\mciteBstWouldAddEndPuncttrue
\mciteSetBstMidEndSepPunct{\mcitedefaultmidpunct}
{\mcitedefaultendpunct}{\mcitedefaultseppunct}\relax
\EndOfBibitem
\bibitem[Moutinho \latin{et~al.}(2021)Moutinho, Eliel, Righi, Gontijo, Paillet,
  Michel, Chiu, Venezuela, and Pimenta]{CVD_TBLG_SciRep}
Moutinho,~M. V.~O.; Eliel,~G. S.~N.; Righi,~A.; Gontijo,~R.~N.; Paillet,~M.;
  Michel,~T.; Chiu,~P.-W.; Venezuela,~P.; Pimenta,~M.~A. Resonance Raman
  enhancement by the intralayer and interlayer electron–phonon processes in
  twisted bilayer graphene. \emph{Scientific Reports} \textbf{2021}, \emph{11},
  17206\relax
\mciteBstWouldAddEndPuncttrue
\mciteSetBstMidEndSepPunct{\mcitedefaultmidpunct}
{\mcitedefaultendpunct}{\mcitedefaultseppunct}\relax
\EndOfBibitem
\bibitem[Eliel \latin{et~al.}(2018)Eliel, Moutinho, Gadelha, Righi, Campos,
  Ribeiro, Chiu, Watanabe, Taniguchi, Puech, Paillet, Michel, Venezuela, and
  Pimenta]{CVD_TBLG_Natcomm}
Eliel,~G. S.~N.; Moutinho,~M. V.~O.; Gadelha,~A.~C.; Righi,~A.; Campos,~L.~C.;
  Ribeiro,~H.~B.; Chiu,~P.-W.; Watanabe,~K.; Taniguchi,~T.; Puech,~P.;
  Paillet,~M.; Michel,~T.; Venezuela,~P.; Pimenta,~M.~A. Intralayer and
  interlayer electron–phonon interactions in twisted graphene
  heterostructures. \emph{Nature Communications} \textbf{2018}, \emph{9},
  1221\relax
\mciteBstWouldAddEndPuncttrue
\mciteSetBstMidEndSepPunct{\mcitedefaultmidpunct}
{\mcitedefaultendpunct}{\mcitedefaultseppunct}\relax
\EndOfBibitem
\bibitem[Campos-Delgado \latin{et~al.}(2013)Campos-Delgado, Cançado, Achete,
  Jorio, and Raskin]{CVD_TBLG_Nanoresearch}
Campos-Delgado,~J.; Cançado,~L.~G.; Achete,~C.~A.; Jorio,~A.; Raskin,~J.-P.
  Raman scattering study of the phonon dispersion in twisted bilayer graphene.
  \emph{Nano Research} \textbf{2013}, \emph{6}, 269--274\relax
\mciteBstWouldAddEndPuncttrue
\mciteSetBstMidEndSepPunct{\mcitedefaultmidpunct}
{\mcitedefaultendpunct}{\mcitedefaultseppunct}\relax
\EndOfBibitem
\bibitem[Huang \latin{et~al.}(2017)Huang, Yankowitz, Chattrakun, Sandhu, and
  LeRoy]{CVD_TBLG_SciRep_doping}
Huang,~S.; Yankowitz,~M.; Chattrakun,~K.; Sandhu,~A.; LeRoy,~B.~J. Evolution of
  the electronic band structure of twisted bilayer graphene upon doping.
  \emph{Scientific Reports} \textbf{2017}, \emph{7}, 7611\relax
\mciteBstWouldAddEndPuncttrue
\mciteSetBstMidEndSepPunct{\mcitedefaultmidpunct}
{\mcitedefaultendpunct}{\mcitedefaultseppunct}\relax
\EndOfBibitem
\bibitem[Schäpers \latin{et~al.}(2022)Schäpers, Sonntag, Valerius, Pestka,
  Strasdas, Watanabe, Taniguchi, Wirtz, Morgenstern, Beschoten, Dolleman, and
  Stampfer]{AS_TBLG_2Dmater}
Schäpers,~A.; Sonntag,~J.; Valerius,~L.; Pestka,~B.; Strasdas,~J.;
  Watanabe,~K.; Taniguchi,~T.; Wirtz,~L.; Morgenstern,~M.; Beschoten,~B.;
  Dolleman,~R.~J.; Stampfer,~C. Raman imaging of twist angle variations in
  twisted bilayer graphene at intermediate angles. \emph{2D Materials}
  \textbf{2022}, \emph{9}, 045009\relax
\mciteBstWouldAddEndPuncttrue
\mciteSetBstMidEndSepPunct{\mcitedefaultmidpunct}
{\mcitedefaultendpunct}{\mcitedefaultseppunct}\relax
\EndOfBibitem
\bibitem[Zhang \latin{et~al.}(2019)Zhang, Zhang, Wang, Zhang, Jiang, Deng,
  Zhang, and Qin]{AS_TBLG_inplane_anisotropy}
Zhang,~X.; Zhang,~R.; Wang,~Y.; Zhang,~Y.; Jiang,~T.; Deng,~C.; Zhang,~X.;
  Qin,~S. In-plane anisotropy in twisted bilayer graphene probed by Raman
  spectroscopy. \emph{Nanotechnology} \textbf{2019}, \emph{30}, 435702\relax
\mciteBstWouldAddEndPuncttrue
\mciteSetBstMidEndSepPunct{\mcitedefaultmidpunct}
{\mcitedefaultendpunct}{\mcitedefaultseppunct}\relax
\EndOfBibitem
\bibitem[Barbosa \latin{et~al.}(2022)Barbosa, Gadelha, Ohlberg, Watanabe,
  Taniguchi, Medeiros-Ribeiro, Jorio, and Campos]{AS_TBLG_nearMA}
Barbosa,~T.~C.; Gadelha,~A.~C.; Ohlberg,~D. A.~A.; Watanabe,~K.; Taniguchi,~T.;
  Medeiros-Ribeiro,~G.; Jorio,~A.; Campos,~L.~C. Raman spectra of twisted
  bilayer graphene close to the magic angle. \emph{2D Materials} \textbf{2022},
  \emph{9}, 025007\relax
\mciteBstWouldAddEndPuncttrue
\mciteSetBstMidEndSepPunct{\mcitedefaultmidpunct}
{\mcitedefaultendpunct}{\mcitedefaultseppunct}\relax
\EndOfBibitem
\bibitem[Pizzocchero \latin{et~al.}(2016)Pizzocchero, Gammelgaard, Jessen,
  Caridad, Wang, Hone, Bøggild, and Booth]{tear_and_stack}
Pizzocchero,~F.; Gammelgaard,~L.; Jessen,~B.~S.; Caridad,~J.~M.; Wang,~L.;
  Hone,~J.; Bøggild,~P.; Booth,~T.~J. The hot pick-up technique for batch
  assembly of van der Waals heterostructures. \emph{Nature Communications}
  \textbf{2016}, \emph{7}, 11894\relax
\mciteBstWouldAddEndPuncttrue
\mciteSetBstMidEndSepPunct{\mcitedefaultmidpunct}
{\mcitedefaultendpunct}{\mcitedefaultseppunct}\relax
\EndOfBibitem
\bibitem[Hicks \latin{et~al.}(2011)Hicks, Sprinkle, Shepperd, Wang, Tejeda,
  Taleb-Ibrahimi, Bertran, Fèvre, de~Heer, Berger, and
  Conrad]{DFT9_symmetry_breaking}
Hicks,~J.; Sprinkle,~M.; Shepperd,~K.; Wang,~F.; Tejeda,~A.;
  Taleb-Ibrahimi,~A.; Bertran,~F.; Fèvre,~P.~L.; de~Heer,~W.~A.; Berger,~C.;
  Conrad,~E.~H. Symmetry breaking in commensurate graphene rotational stacking:
  Comparison of theory and experiment. \emph{Physical Review B} \textbf{2011},
  \emph{83}, 205403\relax
\mciteBstWouldAddEndPuncttrue
\mciteSetBstMidEndSepPunct{\mcitedefaultmidpunct}
{\mcitedefaultendpunct}{\mcitedefaultseppunct}\relax
\EndOfBibitem
\bibitem[Uchida \latin{et~al.}(2014)Uchida, Furuya, Iwata, and
  Oshiyama]{DFT10_atomic_corrugation}
Uchida,~K.; Furuya,~S.; Iwata,~J.-I.; Oshiyama,~A. Atomic corrugation and
  electron localization due to Moiré patterns in twisted bilayer graphenes.
  \emph{Physical Review B} \textbf{2014}, \emph{90}, 155451\relax
\mciteBstWouldAddEndPuncttrue
\mciteSetBstMidEndSepPunct{\mcitedefaultmidpunct}
{\mcitedefaultendpunct}{\mcitedefaultseppunct}\relax
\EndOfBibitem
\bibitem[Luican \latin{et~al.}(2011)Luican, Li, Reina, Kong, Nair, Novoselov,
  Geim, and Andrei]{breakdown_TBLG_DFT11}
Luican,~A.; Li,~G.; Reina,~A.; Kong,~J.; Nair,~R.~R.; Novoselov,~K.~S.;
  Geim,~A.~K.; Andrei,~E.~Y. Single-Layer Behavior and Its Breakdown in Twisted
  Graphene Layers. \emph{Physical Review Letters} \textbf{2011}, \emph{106},
  126802\relax
\mciteBstWouldAddEndPuncttrue
\mciteSetBstMidEndSepPunct{\mcitedefaultmidpunct}
{\mcitedefaultendpunct}{\mcitedefaultseppunct}\relax
\EndOfBibitem
\bibitem[Malard \latin{et~al.}(2009)Malard, Pimenta, Dresselhaus, and
  Dresselhaus]{Dresselhaus_phonondispersion}
Malard,~L.; Pimenta,~M.; Dresselhaus,~G.; Dresselhaus,~M. Raman spectroscopy in
  graphene. \emph{Physics Reports} \textbf{2009}, \emph{473}, 51--87\relax
\mciteBstWouldAddEndPuncttrue
\mciteSetBstMidEndSepPunct{\mcitedefaultmidpunct}
{\mcitedefaultendpunct}{\mcitedefaultseppunct}\relax
\EndOfBibitem
\bibitem[Kresse and Furthmüller(1996)Kresse, and Furthmüller]{DFT7_ab_initio}
Kresse,~G.; Furthmüller,~J. Efficiency of ab-initio total energy calculations
  for metals and semiconductors using a plane-wave basis set.
  \emph{Computational Materials Science} \textbf{1996}, \emph{6}, 15--50\relax
\mciteBstWouldAddEndPuncttrue
\mciteSetBstMidEndSepPunct{\mcitedefaultmidpunct}
{\mcitedefaultendpunct}{\mcitedefaultseppunct}\relax
\EndOfBibitem
\bibitem[Kresse and Furthmüller(1996)Kresse, and Furthmüller]{DFT8_ab_initio}
Kresse,~G.; Furthmüller,~J. Efficient iterative schemes for ab initio
  total-energy calculations using a plane-wave basis set. \emph{Physical Review
  B} \textbf{1996}, \emph{54}, 11169--11186\relax
\mciteBstWouldAddEndPuncttrue
\mciteSetBstMidEndSepPunct{\mcitedefaultmidpunct}
{\mcitedefaultendpunct}{\mcitedefaultseppunct}\relax
\EndOfBibitem
\bibitem[NE \latin{et~al.}(2017)NE, Boujnah, Benyoussef, and
  Kenz]{DFT16_AA_AB_bandstructure}
NE,~M. L.~O.; Boujnah,~M.; Benyoussef,~A.; Kenz,~A.~E. Electronic and
  Electrical Conductivity of AB and AA-Stacked Bilayer Graphene with Tunable
  Layer Separation. \emph{Journal of Superconductivity and Novel Magnetism}
  \textbf{2017}, \emph{30}, 1263--1267\relax
\mciteBstWouldAddEndPuncttrue
\mciteSetBstMidEndSepPunct{\mcitedefaultmidpunct}
{\mcitedefaultendpunct}{\mcitedefaultseppunct}\relax
\EndOfBibitem
\bibitem[Popov(2018)]{DFT17_ramanbands_kinks_spikes}
Popov,~V.~N. Raman bands of twisted bilayer graphene. \emph{Journal of Raman
  Spectroscopy} \textbf{2018}, \emph{49}, 31--35\relax
\mciteBstWouldAddEndPuncttrue
\mciteSetBstMidEndSepPunct{\mcitedefaultmidpunct}
{\mcitedefaultendpunct}{\mcitedefaultseppunct}\relax
\EndOfBibitem
\bibitem[Brihuega \latin{et~al.}(2012)Brihuega, Mallet, González-Herrero,
  de~Laissardière, Ugeda, Magaud, Gómez-Rodríguez, Ynduráin, and
  Veuillen]{DFT14_unraveling_vHs_TBLG}
Brihuega,~I.; Mallet,~P.; González-Herrero,~H.; de~Laissardière,~G.~T.;
  Ugeda,~M.~M.; Magaud,~L.; Gómez-Rodríguez,~J.~M.; Ynduráin,~F.;
  Veuillen,~J.-Y. Unraveling the Intrinsic and Robust Nature of van Hove
  Singularities in Twisted Bilayer Graphene by Scanning Tunneling Microscopy
  and Theoretical Analysis. \emph{Physical Review Letters} \textbf{2012},
  \emph{109}, 196802\relax
\mciteBstWouldAddEndPuncttrue
\mciteSetBstMidEndSepPunct{\mcitedefaultmidpunct}
{\mcitedefaultendpunct}{\mcitedefaultseppunct}\relax
\EndOfBibitem
\bibitem[Popov(2015)]{Raman_2phononbands}
Popov,~V.~N. Two-phonon Raman bands of bilayer graphene: Revisited.
  \emph{Carbon} \textbf{2015}, \emph{91}, 436--444\relax
\mciteBstWouldAddEndPuncttrue
\mciteSetBstMidEndSepPunct{\mcitedefaultmidpunct}
{\mcitedefaultendpunct}{\mcitedefaultseppunct}\relax
\EndOfBibitem
\bibitem[Samsonidze \latin{et~al.}(2003)Samsonidze, Saito, Jorio, Filho,
  Grüneis, Pimenta, Dresselhaus, and Dresselhaus]{CNT_1}
Samsonidze,~G.~G.; Saito,~R.; Jorio,~A.; Filho,~A. G.~S.; Grüneis,~A.;
  Pimenta,~M.~A.; Dresselhaus,~G.; Dresselhaus,~M.~S. Phonon Trigonal Warping
  Effect in Graphite and Carbon Nanotubes. \emph{Physical Review Letters}
  \textbf{2003}, \emph{90}, 027403\relax
\mciteBstWouldAddEndPuncttrue
\mciteSetBstMidEndSepPunct{\mcitedefaultmidpunct}
{\mcitedefaultendpunct}{\mcitedefaultseppunct}\relax
\EndOfBibitem
\bibitem[Jorio \latin{et~al.}(2001)Jorio, Saito, Hafner, Lieber, Hunter,
  McClure, Dresselhaus, and Dresselhaus]{CNT_2}
Jorio,~A.; Saito,~R.; Hafner,~J.~H.; Lieber,~C.~M.; Hunter,~M.; McClure,~T.;
  Dresselhaus,~G.; Dresselhaus,~M.~S. Structural (n,m) Determination of
  Isolated Single-Wall Carbon Nanotubes by Resonant Raman Scattering.
  \emph{Physical Review Letters} \textbf{2001}, \emph{86}, 1118--1121\relax
\mciteBstWouldAddEndPuncttrue
\mciteSetBstMidEndSepPunct{\mcitedefaultmidpunct}
{\mcitedefaultendpunct}{\mcitedefaultseppunct}\relax
\EndOfBibitem
\bibitem[Cong \latin{et~al.}(2011)Cong, Yu, Saito, Dresselhaus, and
  Dresselhaus]{combination_modes}
Cong,~C.; Yu,~T.; Saito,~R.; Dresselhaus,~G.~F.; Dresselhaus,~M.~S.
  Second-Order Overtone and Combination Raman Modes of Graphene Layers in the
  Range of 1690 - 2150 cm$^{-1}$. \emph{ACS Nano} \textbf{2011}, \emph{5},
  1600--1605\relax
\mciteBstWouldAddEndPuncttrue
\mciteSetBstMidEndSepPunct{\mcitedefaultmidpunct}
{\mcitedefaultendpunct}{\mcitedefaultseppunct}\relax
\EndOfBibitem
\bibitem[Rao \latin{et~al.}(2011)Rao, Podila, Tsuchikawa, Katoch, Tishler, Rao,
  and Ishigami]{layer_stacking_combinationmodes}
Rao,~R.; Podila,~R.; Tsuchikawa,~R.; Katoch,~J.; Tishler,~D.; Rao,~A.~M.;
  Ishigami,~M. Effects of Layer Stacking on the Combination Raman Modes in
  Graphene. \emph{ACS Nano} \textbf{2011}, \emph{5}, 1594--1599\relax
\mciteBstWouldAddEndPuncttrue
\mciteSetBstMidEndSepPunct{\mcitedefaultmidpunct}
{\mcitedefaultendpunct}{\mcitedefaultseppunct}\relax
\EndOfBibitem
\bibitem[Woods \latin{et~al.}(2014)Woods, Britnell, Eckmann, Ma, Lu, Guo, Lin,
  Yu, Cao, Gorbachev, Kretinin, Park, Ponomarenko, Katsnelson, Gornostyrev,
  Watanabe, Taniguchi, Casiraghi, Gao, Geim, and Novoselov]{graphene_hBN}
Woods,~C.~R. \latin{et~al.}  Commensurate–incommensurate transition in
  graphene on hexagonal boron nitride. \emph{Nature Physics} \textbf{2014},
  \emph{10}, 451--456\relax
\mciteBstWouldAddEndPuncttrue
\mciteSetBstMidEndSepPunct{\mcitedefaultmidpunct}
{\mcitedefaultendpunct}{\mcitedefaultseppunct}\relax
\EndOfBibitem
\bibitem[Kresse and Joubert(1999)Kresse, and Joubert]{DFT5_psuedopotential}
Kresse,~G.; Joubert,~D. From ultrasoft pseudopotentials to the projector
  augmented-wave method. \emph{Physical Review B} \textbf{1999}, \emph{59},
  1758--1775\relax
\mciteBstWouldAddEndPuncttrue
\mciteSetBstMidEndSepPunct{\mcitedefaultmidpunct}
{\mcitedefaultendpunct}{\mcitedefaultseppunct}\relax
\EndOfBibitem
\bibitem[Blöchl(1994)]{DFT6_augmentedwave}
Blöchl,~P.~E. Projector augmented-wave method. \emph{Physical Review B}
  \textbf{1994}, \emph{50}, 17953--17979\relax
\mciteBstWouldAddEndPuncttrue
\mciteSetBstMidEndSepPunct{\mcitedefaultmidpunct}
{\mcitedefaultendpunct}{\mcitedefaultseppunct}\relax
\EndOfBibitem
\bibitem[Perdew \latin{et~al.}(1996)Perdew, Burke, and
  Ernzerhof]{DFT4_gradient_approach}
Perdew,~J.~P.; Burke,~K.; Ernzerhof,~M. Generalized Gradient Approximation Made
  Simple. \emph{Physical Review Letters} \textbf{1996}, \emph{77},
  3865--3868\relax
\mciteBstWouldAddEndPuncttrue
\mciteSetBstMidEndSepPunct{\mcitedefaultmidpunct}
{\mcitedefaultendpunct}{\mcitedefaultseppunct}\relax
\EndOfBibitem
\bibitem[Grimme(2006)]{DFT3_GGA}
Grimme,~S. Semiempirical GGA-type density functional constructed with a
  long-range dispersion correction. \emph{Journal of Computational Chemistry}
  \textbf{2006}, \emph{27}, 1787--1799\relax
\mciteBstWouldAddEndPuncttrue
\mciteSetBstMidEndSepPunct{\mcitedefaultmidpunct}
{\mcitedefaultendpunct}{\mcitedefaultseppunct}\relax
\EndOfBibitem
\bibitem[Monkhorst and Pack(1976)Monkhorst, and Pack]{DFT2_special_points}
Monkhorst,~H.~J.; Pack,~J.~D. Special points for Brillouin-zone integrations.
  \emph{Physical Review B} \textbf{1976}, \emph{13}, 5188--5192\relax
\mciteBstWouldAddEndPuncttrue
\mciteSetBstMidEndSepPunct{\mcitedefaultmidpunct}
{\mcitedefaultendpunct}{\mcitedefaultseppunct}\relax
\EndOfBibitem
\end{mcitethebibliography}

\end{document}